\documentclass[Preprint, 12pt, authoryear]{elsarticle}

\usepackage{amssymb}
\usepackage[T1]{fontenc}
\usepackage{float}
\usepackage{amsmath} 
\usepackage{cleveref}
\usepackage[table]{xcolor}
\usepackage{multirow}
\usepackage{subcaption}
\usepackage{url}
\usepackage{enumitem}
\definecolor{subrowcolor1}{RGB}{240,240,240}
\definecolor{subrowcolor2}{RGB}{220,220,220}
\definecolor{subrowcolor3}{RGB}{200,200,200}

\usepackage{algorithm}
\usepackage[noend]{algpseudocode}

\newcommand{\rev}[2][black]{\textcolor{#1}{#2}}
\newcommand{\revt}[2][black]{\textcolor{#1}{#2}}

\DeclareMathOperator*{\argmax}{arg\,max}
\journal{Computational Materials Science}

\begin{document}
	
	\begin{frontmatter}
		
		\title{A Novel Constrained Sampling Method for Efficient Exploration in Materials and Chemical Mixture Design\textsuperscript{**}}

		\author[inst1]{Christina Schenk\corref{cor1}}
		\ead{christina.schenk@imdea.org}
		\cortext[cor1]{Corresponding author}
		
		\affiliation[inst1]{organization={IMDEA Materials Institute},
			addressline={Eric Kandel 2, Tecnogetafe}, 
			city={Getafe},
			postcode={28906}, 
			state={Madrid},
			country={Spain}}

		\author[inst1]{Maciej Haranczyk}
		\begin{abstract}	
			Efficient exploration of multicomponent material composition spaces is often limited by time and financial constraints, particularly when mixture and synthesis constraints exist. Traditional methods like Latin hypercube sampling (LHS) struggle with \rev{constrained problems especially in high dimensions}\revt{,} while emerging approaches like Bayesian optimization (BO) face challenges in early-stage exploration. \rev{This article introduces ConstrAined Sequential laTin hypeRcube sampling methOd (CASTRO), an open-source tool designed to address these challenges.} \revt{CASTRO is optimized for uniform sampling in constrained small- to moderate-dimensional spaces, with scalability to higher dimensions through future adaptations. CASTRO uses a divide-and-conquer strategy to decompose problems into parallel subproblems, improving efficiency and scalability.} It effectively handles equality-mixture constraints, ensuring comprehensive design space coverage and leveraging LHS and LHS with multidimensional uniformity (LHSMDU). \revt{It also integrates prior experimental knowledge, making it well-suited for efficient exploration within limited budgets. Validation through two material design case studies, a four-dimensional problem with near-uniform distributions and a nine-dimensional problem with additional synthesis constraints, demonstrates CASTRO's effectiveness in exploring constrained design spaces for materials science, pharmaceuticals and chemicals.} The software and case studies are available on GitHub.\end{abstract}
		
		
		
		
	\end{frontmatter}
	\noindent\textsuperscript{**}Accepted version. Accepted for publication in Computational Materials Science on February 10, 2025.
	
	\section{Introduction}
	For many engineering applications, the design of experiments plays a crucial role. Although traditional approaches, such as quasi-random search sampling methods such as Latin hypercube sampling (LHS), continue to be widely used, contemporary investigations are increasingly focusing on adaptive experimentation through Bayesian optimization (BO). This shift aims to achieve autonomous experimental setups and high-throughput pipelines, making experimentation more efficient and cost-effective.
	
	This shift is particularly relevant in materials science, where discovering novel chemicals and materials requires optimizing \revt{specific properties such as thermal, mechanical, or optical performance}. Machine learning (ML) models have become powerful tools in this space, enabling researchers to predict material behaviors and navigate complex design spaces \rev{\citep{STERGIOU2023112031}}. However, even with the aid of ML-driven optimization, the challenge of designing constrained \revt{experiments, where factors such as mixture or volume constraints limit the feasible space, persists}.
	
	Constrained experimental design plays a pivotal role in various fields, particularly in materials science, where mixture and volume constraints often govern experimental setups. Examples include the design of glass compositions \citep{Borbowski2009}, pharmaceutical formulations \citep{Cafaggi2003}, rheological clay–polymer compositions \citep{LoDico2022} and chemical compositions in food science \citep{Kpodo2013}. \rev{Conventional approaches like LHS can struggle to maintain uniformity in high-dimensional constrained spaces due to the challenge of confining samples to lower-dimensional manifolds (e.g., simplices)\revt{\citep{fang2005, Wang2019}}. In low dimensions, these deficiencies can be mitigated by incorporating constraints directly into the sampling method, such as through normalization or projection techniques \revt{\citep{santner2003design, fang2005}}. However, in medium to high-dimensional \revt{constrained} spaces, while normalization and projection help to enforce constraints, they do not fully resolve the uniformity and space-filling issues due to the curse of dimensionality and the concentration of measure, a phenomenon where \revt{random} points in high-dimensional spaces tend to cluster near certain values (e.g., the mean or expected value) as dimensionality increases \revt{\citep{Esposito2023}}. These effects lead to uneven exploration of the space \revt{\citep{santner2003design}}. One way to address these deficiencies is to use additional sampling methods, such as Dirichlet sampling \revt{\citep{gelman2013bayesian}} or modified space-filling designs \revt{\citep{MORRIS1995}}, which are specifically tailored to improve the uniformity and coverage of the constrained space.} 
	
	\rev{Several distance-based strategies, such as maximin or minimax designs, exist for generating robust, uniform, and well-distributed sampling points \citep{JOHNSON1990}. Additionally, exploratory designs aim to balance criteria like entropy or maximin while ensuring good projective properties in each dimension \citep{MORRIS1995}. 
		\citet{Joseph2016} provides a review for space-filling designs including minimax and maximin distance designs and maximum projection designs \citep{joseph_maximum_2015}.
		However, these methods typically do not inherently account for constraints in their traditional form. Adaptations or extensions are required to handle constraints, such as modifying the optimization problem or filtering samples to ensure feasibility. Moreover, the projections onto the subspaces with dimensions $2,\ldots n-1$ may not always exhibit good coverage \citep{Joseph2016}.
		Recently, some efforts have introduced improved distance-based criteria for Latin hypercube sampling and other methods by incorporating periodic distance metrics \citep{vorechovsky2020}.}
	
	While machine learning and optimization strategies, such as BO, can assist in navigating these spaces, they are often reliant on surrogate models and require a significant number of initial experiments to become reliable. The early stages of adaptive experimentation often prioritize pure exploration to improve the surrogate model, but there is no guarantee that this exploration will adequately cover the entire design space. \rev{Achieving uniformity under mixture and equality constraints remains challenging for standard LHS \revt{\citep{mckay1979comparison}} because it does not guarantee joint stratification within the constrained region. This issue can be illustrated through distribution analyses as shown in Additional Figures in Supplementary Material \citep{supplementary}, where gaps or clustering often appear compared to methods specifically designed for simplices (e.g., Dirichlet sampling).}
	
	Our constrained design of experiments (DOE) approach directly addresses these challenges by offering a methodology designed to generate uniform and space-filling designs in constrained spaces \rev{for small-to-medium-dimensional problems although technically not limited to the latter.} This is achieved through novel sampling techniques that ensure efficient exploration of the experimental design space while respecting the imposed constraints. Unlike standard approaches, such as point distance-based optimization methods, our approach focuses specifically on maintaining uniformity in constrained regions, a critical feature often overlooked by traditional techniques \citep{schneider_latin_2023}.
	
	Several existing methods attempt to tackle constrained DOE problems. \citet{petelet_latin_2010} introduced a methodology for Latin hypercube sampling with inequality constraints. \citet{Borbowski2009} proposed two number-theoretic methods for building space-filling and in particular uniform designs for constrained mixture experiments involving single and multiple-component constraints. \citet{liu_construction_2015} developed a new method based on the central composite discrepancy criterion for irregular regions and the switching algorithm from \citet{chuang_uniform_2010}. More recently, \citet{jourdan_space-filling_2023} utilized an optimization method to build mixture experimental designs targeting a Dirichlet distribution.
	While these approaches have made strides in constrained DOE, challenges remain in particular in high-dimensional spaces. \citet{schneider_latin_2023,schneider_uniformdesign2023} introduced a projection-based method that maps uniformly distributed designs to the constraint using incremental Latin hypercube sampling \citep{Voigt2020,schneider_latin_2023}, slack variable concepts, and maximin Latin hypercubes \citep{schneider_uniformdesign2023}. These methodologies offer alternatives to permutation-based approaches by employing optimization strategies. Despite claims of limited impact from the curse of dimensionality, the latest developed method by Schneider et al. exhibits certain limitations, particularly in cases where constraints lack a unique feasible solution for projecting the support design onto the constraint.
	Additionally, \citet{LIU2019285} used an optimization-based method involving mixed-integer nonlinear programming to design molecular mixtures. While these methods can handle different variable types, solving such problems can become computationally expensive.
	
	In adaptive experimentation, particularly with BO, integrating constraints can take various forms, but it often introduces challenges that render the problems ill-posed. Moreover, these methods heavily rely on surrogate models, demanding specific computational setups and modifications. At the beginning of adaptive experimentation, the surrogate model may lack reliability. Nevertheless, this approach facilitates the generation of more points adaptively. Typically, the early stages of the optimization process in adaptive experimentation are dedicated to pure exploration. This phase usually involves a fixed number of steps determined by the degrees of freedom. However, there is no assurance that the samples are evenly distributed throughout the entire design space across all dimensions. Although traditional sampling methods such as LHS, Sobol, Halton, and Hammersley can predefine the number of random points to sample in a space-filling manner, doing so in \rev{high-dimensional constrained spaces} is not straightforward and demands specialized methods.
	
	Several works have shown that BO, particularly when integrated with machine learning-driven acquisition functions on average can be more efficient. However, depending on the landscape of the problem, that is, whether there are multiple optima and where they are located, performing a pure exploration phase before moving to other acquisition functions to balance exploration and exploitation or pure exploitation can be important \citep{DeAth2021}. This can be specifically relevant if experiments are costly and we want to minimize the number of experiments executed for exploration to get a reliable surrogate model. 
	
	LHS and the BO pure exploration strategy have shown comparable performance in several test problems \citep{DeAth2021}. However, for complex landscapes, due to the space-filling property, one could assume that choosing a quasi-random search sampling method for the exploration phase may be beneficial to get a better initial surrogate model with fewer required samples compared to adaptive experimentation.
	
	To assess space-filling properties and statistically quantify uniformity, researchers often rely on various discrepancy measures. Common metrics include $L_{\infty}$-star discrepancy, $L_2$-star discrepancy, centered $L_2$-discrepancy, and wrap-around $L_2$ discrepancy \citep{zhou_mixture_2013}. Of these, the centered $L_2$-discrepancy and wrap-around $L_2$ discrepancy are particularly important in experimental design, as they satisfy all relevant criteria for evaluating uniformity, as outlined by \citet{fang2005}. In irregular regions, such as those imposed by mixture constraints, several other widely used discrepancy measures exist \citep{liu_construction_2015}. These include the mean squared error (MSE), root mean squared distance (RMSD), maximum distance (MD), average distance (AD) discrepancies \citep{Borbowski2009}, and the central composite discrepancy (CCD) \citep{chuang_uniform_2010}. Each of these measures provides valuable insights into the distribution and uniformity of samples in experimental designs.
	
	To address the challenges posed by constrained \rev{high-dimensional spaces}, we propose a novel sampling strategy that ensures uniformity and space-filling properties under mixture and other constraints. \revt{Our method provides a flexible and efficient alternative for experimental design, combining advanced sampling techniques with the ability to handle complex constraints across a range of dimensionalities. While optimized for small- to moderate-dimensional problems, the method is inherently scalable. Its divide-and-conquer approach decomposes problems into subproblems that can be sampled in parallel, improving efficiency. Through future adaptations, this approach can be extended to high-dimensional spaces, helping to mitigate some of the challenges associated with the curse of dimensionality.} Additionally, we maximize the use of existing expensive experimental data by strategically incorporating new experiments to fill gaps in the design space. This hybrid approach allows researchers to adhere to budget constraints while maximizing exploration in constrained experimental landscapes. We evaluate the space-filling properties of our approach by analyzing both the centered and wrap-around $L_2$ discrepancies, along with the variance of the samples. These metrics are then compared to those obtained from \rev{scaled} traditional DOE methods, providing a theoretical baseline for comparison. \revt{ In addition, we perform distribution analysis to assess how well the generated samples represent the target design space, ensuring comprehensive coverage and complementing the previously collected data under the imposed constraints.}
	
	In \Cref{sec:meth}, we introduce the novel methodology for identifying the experiments to be carried out. This includes the division of the original problem into subproblems, an explanation of the space-filling constrained sampling, and the required post-processing steps. Moving to \Cref{sec:res}, we apply these methods to two practical problems within materials science. Here, we analyze the results focusing on uniformity and the space-filling property. Finally, we conclude with a summary of the main findings in \Cref{sec:con}.
	
	\section{Methods}\label{sec:meth}
	\subsection{Challenges in Experimental Design for Chemists}
	\rev{While modern Design of Experiments (DOE) techniques such as factorial designs, response surface methodology (RSM), and advanced optimization methods like Nelder-Mead, genetic algorithms, and Bayesian optimization have revolutionized experimental design, there are still cases where chemists rely on traditional methods. In some situations, experimental data is still collected based on the chemist's knowledge and experience, using expensive testing procedures that require significant time and resources. These experimental procedures can be resource-intensive. However, the integration of advanced DOE methods and computational tools has significantly enhanced the efficiency, cost-effectiveness, and ability to handle complex, high-dimensional data. These developments highlight the importance of combining chemical expertise with cutting-edge optimization strategies to further improve the overall experimental process.}
	
	\rev{When seeking computational support to explore the design space or statistically relevant compositions, chemists often find that they have already conducted several costly experiments. To minimize efforts and costs, a methodology that can incorporate preliminary data while handling mixture and synthesis constraints is highly advantageous. Such a method would reduce the number of additional experiments needed to identify promising compositions by taking previously collected data into account. The method presented in the remaining parts of this section addresses these needs.}
	\subsection{Algorithmic Details}
\begin{figure}[H]
		\centering
		\includegraphics[scale=.5]{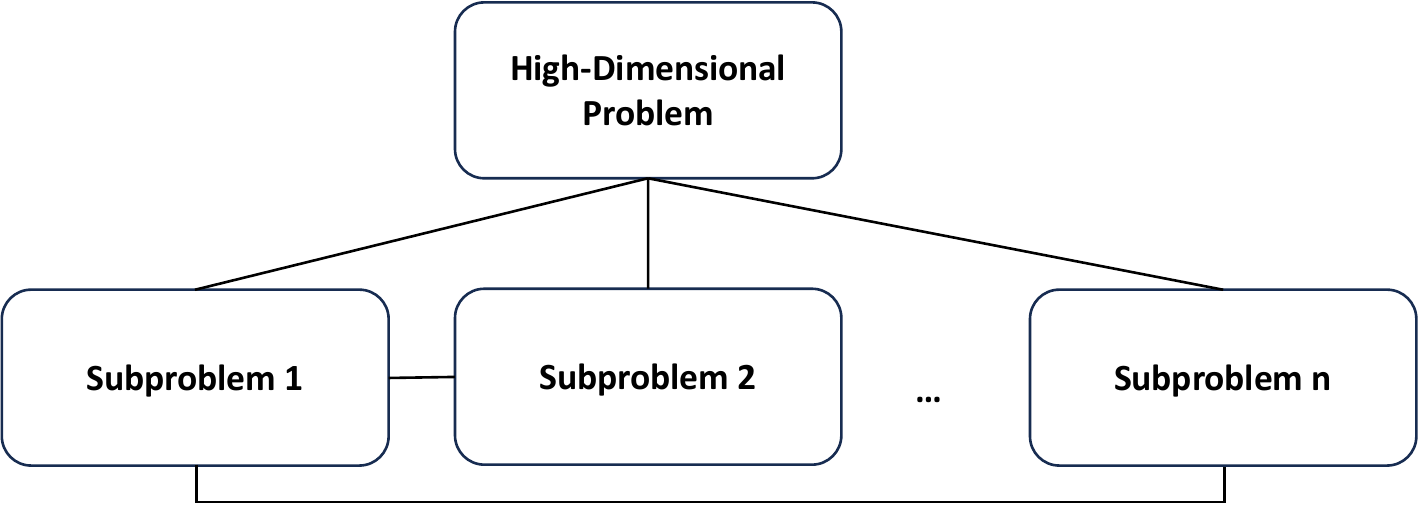}
		\caption{Overview of the division of the \rev{full-dimensional} problem into n subproblems}
		\label{fig:overvhighdimprob}
	\end{figure}

	In the following, we introduce an algorithm for experimental design that handles equality constraints, such as ensuring fractions sum to one. \rev{While our method is effective for up to four dimensions due to the curse of dimensionality, this is not a strict limitation. In fact, we have implemented a divide-and-conquer strategy to address higher-dimensional (>4) problems. This approach divides the original higher-dimensional problem (>4) into a main problem and multiple lower-dimensional subproblems, as shown in \Cref{fig:overvhighdimprob}. Each subproblem is solved individually, and the results are then integrated back into the full-dimensional solution. Specifically, the experimental data are rescaled based on the division of the original problem, ensuring that fractions sum to one for each subproblem.}
	\revt{For the scope of this work, we focus on concentrations but modern accelerated discovery and optimization platforms demand the control of additional parameters such as time, temperature, and pH. These could be categorized into separate subproblems, allowing the methodology to be extended to handle such factors as part of the overall experimental design.}
	\rev{While the examples presented in this paper primarily focus on small- to medium-dimensional problems, technically the method is not confined to these and can be extended for higher-dimensional cases as well.}
	
	\rev{The algorithm can be executed deterministically, i.e. for just one seed or it can be executed multiple times with different random seeds, and then the results can be combined and then the most distant samples leading to overall uniformity can be selected. In the examples presented in this paper, we focus on the stochastic version.}
	
	After obtaining the CASTRO suggestions for each subproblem, we select $n_{exp} + {des}_{n_{samp}}$ points that are the farthest from the experimental data based on their Euclidean distances, where $n_{exp}$ is the number of previously collected data points and ${des}_{n_{samp}}$ is the number of desired experiments. We then reassemble the suggestions for each subproblem to obtain the final recommendations for the original problem. This involves selecting the ${des}_{n_{samp}}$ most distant points for the main problem by calculating Euclidean distances and the ${des}_{n_{samp}}$ random points for problems with synthesis constraints. The samples are rescaled so that fractions sum to one for the entire problem. The specifics of preprocessing and postprocessing for different examples are detailed in \Cref{sec:res}. 
	
	\rev{Here, we utilize the Euclidean distance. However, \citet{vorechovsky_distance-based_2019,vorechovsky_distance-based_2020} highlight that in high-dimensional design spaces, using Euclidean distance can lead to a concentration of points around the mean value which remains an important consideration when applying the strategy presented here to high-dimensional cases.}
	
\begin{figure}[H]
		\centering
		\includegraphics[scale=.5]{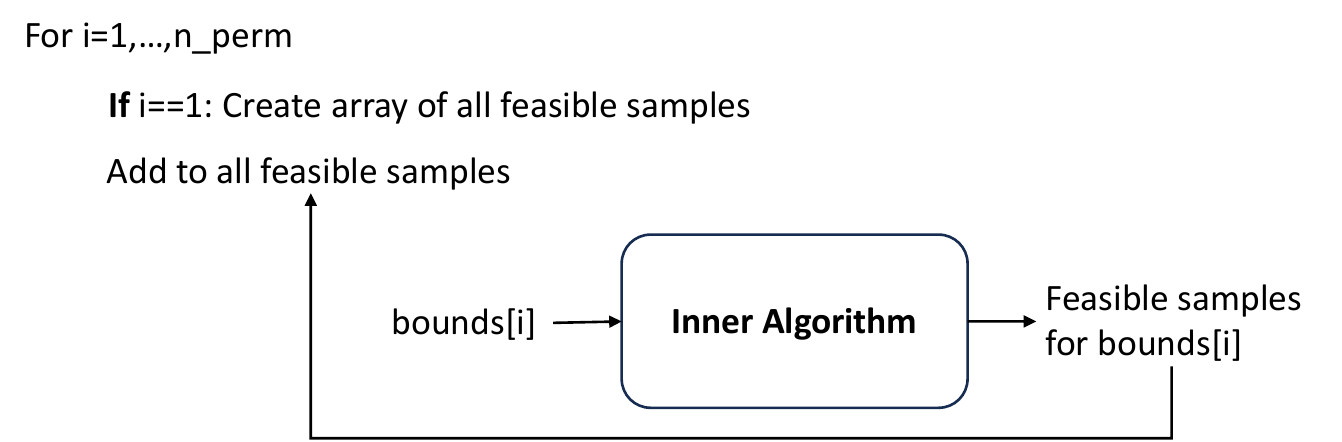}
		\caption{Overview of algorithm: Connection of outer and inner algorithm}
		\label{fig:outeralg}
	\end{figure}

\begin{figure}[H]
		\centering
		\includegraphics[scale=.48]{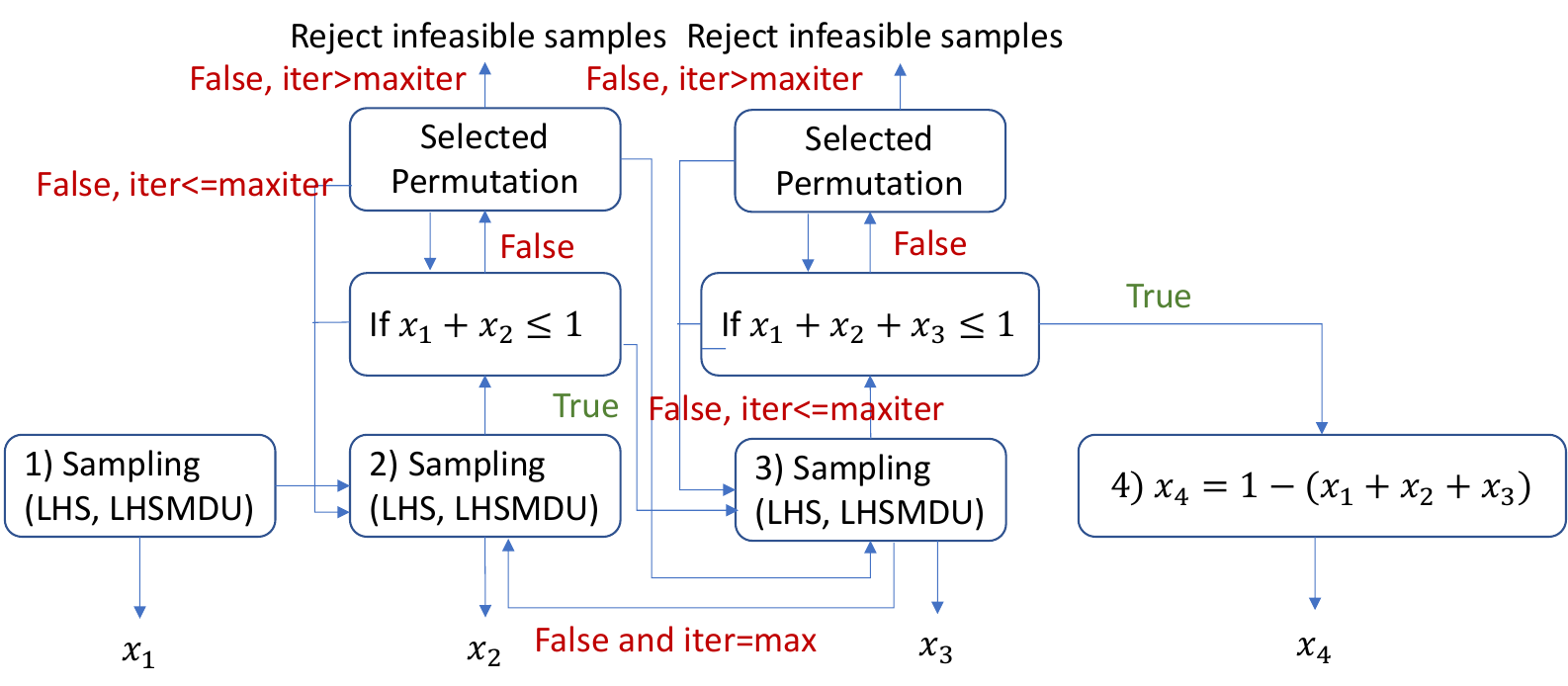}
		\caption{Overview of basic concepts of inner algorithm}
		\label{fig:inneralg}
	\end{figure}

	Next, we will focus on the algorithmic aspects.
	To ensure that the selected samples are equally distributed and independent of the order of the bounds (where typically smaller values are chosen last to meet constraints), we incorporate an outer algorithm that calls an inner algorithm multiple times -- equal to the number of permutations of the bounds. The outer and inner algorithms are connected as visualized in \Cref{fig:outeralg}. The outer algorithm iterates over all permutations of the bounds, running the inner algorithm for each permutation. The pseudo-code for this algorithm is provided in \cref{alg:boundperm}. Feasible samples from each permutation are added to the collection of all feasible samples in the order of the first permutation. The basic idea of the inner algorithm is depicted in \Cref{fig:inneralg}. Depending on the dimensionality, different versions of the algorithm are executed, as detailed in Algorithms 1, 2, 3, 5 in Supplementary Material \citep{supplementary}, with the latter two involving corresponding permutation subalgorithms in Algorithm 4 and Algorithm 6 in Supplementary Material \citep{supplementary}. 
	\begin{algorithm}
		\caption{Inner Algorithm}
		\begin{algorithmic}[1]
			\State \textbf{Variables and Parameters:}
			\State $\textit{dim}$: Dimension of the problem.
			\State $n_{samp}$: The number of samples depends on $tot_{samp}$ typically chosen such that divides exactly the number of all bound permutations ($all_{perms}$) and is larger than $n_{exp}+des_{n_{samp}}$ with
			$n_{samp} = tot_{samp}//len(all_{perms})$.
			\State $max_{rej}$: Maximum number of rejections allowed.
			\State $max_{{iter}_{dim2}}$: Maximum iterations allowed for dimension 2.
			\State $l_1, l_2$: Counting indices for iterations.
			
			\Statex
			\State Determine the dimension (\textit{dim}) of the problem, the number of samples $n_{samp}$, the maximum number of rejections allowed $max_{rej}$ and if dim >2 the maximum number of iterations allowed for dimension 2 $max_{{iter}_{dim2}}$.
			\State Based on the dimension, select the appropriate algorithm (Algorithm 1, 2, or 3) as described in the Supplementary Material (Schenk and Haranczyk, 2024).
			\State Initialize a counting index $l_1$.
			
			\If{$\text{dim} > 2$}
			\State Calculate \textit{sample1} and \textit{sample2}.
			\State Permute these samples until the number of feasible samples is greater than or equal to $n_{\text{samp}} - max_{rej}$ or until $l_1$ exceeds $max_{{iter}_{dim2}}$.
			\If{necessary}
			\State Choose a different permutation strategy using Algorithm 4 (increase the counter).
			\EndIf
			\EndIf
			
			\If{$\text{dim} > 3$}
			\State Calculate \textit{sample3}.
			\State Initialize another counting index $l_2$.
			\State Permute samples using Algorithm 5, similar to the previous step for $\text{dim} = 3$, and increase the counter.
			\EndIf
			
			\State Calculate the last component using $1 - \sum_i^{\text{dim}} \textit{sample}_i$.
			\State Perform an additional bound check on the calculated component.
			
			\If{bounds are fulfilled}
			\State \textbf{Stop the algorithm.}
			\Else
			\State Remove the samples that do not meet the bounds.
			\State \textbf{Stop the algorithm.}
			\EndIf
			
		\end{algorithmic}
	\end{algorithm}
	
	\begin{algorithm}
		\caption{Bound Permutation Algorithm}\label{alg:boundperm}
		
		\begin{algorithmic}[1]
			\State \textbf{Variables and Parameters:}
			\State $num_{\text{meth}}$: Counter for the number of methods tried (0 or 1).
			\State $perm_{\text{ind}}$: Index of the current permutation.
			\State $\text{combi}$: Combination of the current permutation.
			\State $all_{\text{perms}}$: List of all possible permutations.
			\State $bounds$: Bounds for the current dimension.
			\State $methodname$: Name of the current method ("LHS" or "LHSMDU").
			\State $samples$: Samples generated by the Conditioned Sampling Algorithm.
			\State $all\_val\_samples$: Stack of all valid samples.
			\State $all\_val\_samples\_mdu$: Stack of all valid samples for the LHSMDU method.
			\State $val\_samples\_unord$: Valid unordered samples.
			\State $val\_samples\_ord$: Valid ordered samples.
			\State $all\_select$: Flag to determine if all valid samples should be selected.
			\algstore{boundpermstate}
		\end{algorithmic}
	\end{algorithm}
	\begin{algorithm}
		\caption*{Bound Permutation Algorithm (continued)}
		\begin{algorithmic}[1]
			\algrestore{boundpermstate}
			\State $num_{\text{meth}} \gets 0$
			\While{$num_{\text{meth}} < 2$}
			\For{$perm_{\text{ind}}, \text{combi}$ in $\text{enumerate}(all_{\text{perms}})$}
			\State $bounds \gets \text{get\_bounds\_for\_dimension}$ 
			\State $methodname \gets \text{"LHS"}$ if $num_{\text{meth}} == 0$ else $\text{"LHSMDU"}$
			\State $samples \gets \text{Conditioned Samples from Algorithm}~\ref{fig:inneralg}$
			
			\If{$perm_{\text{ind}} == 0$}
			\If{$num_{\text{meth}} == 0$}
			\State $all\_val\_samples \gets \text{stack\_samples}(samples, \text{dim})$
			\Else
			\State $all\_val\_samples\_mdu \gets \text{stack\_samples}(samples, \text{dim})$
			\EndIf
			\Else
			\State $val\_samples\_unord \gets \text{stack\_samples}(samples, \text{dim})$
			\State $val\_samples\_ord \gets \text{np.zeros\_like}(val\_samples\_unord)$
			
			\For{$num, \text{ind}$ in $\text{enumerate}(\text{combi})$}
			\State $val\_samples\_ord[:, \text{ind}] \gets val\_samples\_unord[:, num]$
			\EndFor
			
			\If{all\_select}
			\State $all\_val\_samples \gets \text{np.vstack}((all\_val\_samples, val\_samples\_ord))$
			\Else
			\State $all\_val\_samples \gets \text{select samples by checking distance}$
			\EndIf
			\EndIf
			\EndFor
			
			\State $num_{\text{meth}} \gets num_{\text{meth}} + 1$
			
			\If{$num_{\text{meth}} < 2$}
			\If{$num_{\text{meth}} == 1$}
			\State $all\_val\_samples\_0 \gets all\_val\_samples$
			\Else
			\State \Return $all\_val\_samples\_0, all\_val\_samples$
			\EndIf
			\EndIf
			\EndWhile
		\end{algorithmic}
	\end{algorithm}
	
	\begin{algorithm}
		\caption{Data Distance Check Algorithm}
		\label{alg:datadistcheck}
		\begin{algorithmic}[1]
			
			\State \textbf{Variables and Parameters:}
			\State $samples\_LHS$: Samples generated using Conditioned Latin Hypercube Sampling (LHS) method.
			\State $samples\_LHSMDU$: Samples generated using the Conditioned LHS with Multi-Dimensional Uniformity (LHSMDU) method.
			\State $des_{n_{samp}}$: Desired number of samples to select.
			\State $tol\_samples$: Selected samples from samples\_LHS.
			\State $tol\_samples\_LHSMDU$: Selected samples from samples\_LHSMDU.
			
			\Statex
			\Statex
			\Require Experimental data, $samples\_LHS$, $samples\_LHSMDU$
			\Ensure $tol\_samples$, $tol\_samples\_LHSMDU$

			\State Calculate Euclidean distances:
			\State \quad From $samples\_LHS$/$samples\_LHSMDU$ to experimental data
			\State \quad Among $samples\_LHS$/$samples\_LHSMDU$ themselves
			
			\State Select $des_{n_{samp}}$ samples:
			\State \quad from $samples\_LHS$ with maximum distance to experimental data
			\State \quad from $samples\_LHSMDU$ with maximum distance to experimental data
			\State \quad Round selected samples to desired decimals    
			\State \Return $tol\_samples$, $tol\_samples\_LHSMDU$
		\end{algorithmic}
	\end{algorithm}
	
	In the inner algorithm, we generate $n_{samp}$ samples for each component and permutation. Samples are collected sequentially for each component, with checks to ensure that they sum to one. We track valid combinations using a matrix that records the sum values for combinations (i,j) and by adding and removing the pairs from index lists. If we cannot find a feasible combination after $max_{iter}$ iterations but have found $n_{samp} - \max_{rej}$ feasible samples, we stop. Otherwise, we randomly select a feasible pair for the missing index from the feasible tuple index list and check if the second index is already among the feasible samples found. If not, we add this pair; if so, we remove the corresponding pair and continue. After calculating the last component via $1-\sum_{i=1}^{3}sample_i$, an additional check ensures that the sample satisfies the bounds of this component.
	Several configurations of the algorithm can be adjusted. The user can choose between standard Latin Hypercube Sampling using the \emph{scipy.stats.qmc.lhs} module or Latin Hypercube Sampling with multidimensional uniformity capitalizing the \emph{lhsmdu} Python package. Additionally, there is an option to select $num_{select}$ feasible samples with the greatest Euclidean distance from already selected feasible samples or to select all samples. This option is controlled by setting $all\_select=False$ and specifying $num\_select$.  
	
	Once all feasible samples are found, we choose $des_{n_{samp}}+n_{exp}$ points based on their Euclidean distance from previously collected experimental data for all subproblems. Then we reassemble the problem, ensuring that the fractions sum to one. Depending on the capacity of the experiments that can be performed, that is, $des_{n_{samp}}$, we choose $des_{n_{samp}}$ samples that have the greatest Euclidean distance from those previously collected and ensure a minimum distance between each other. The pseudo-code for this process is outlined in \cref{alg:datadistcheck}.
	\subsection{Distribution and Installation}\label{sect:dist}
	CASTRO leverages a variety of common Python packages for data processing, including \emph{numpy}, \emph{scipy}, \emph{pandas}, \emph{random}, \emph{scikit-learn}, \emph{sympy} and \emph{itertools}. For LHS and LHSMDU sampling, it uses the lhs sampling function from \emph{scipy.stats.qmc} and the \emph{lhsmdu} package. In postprocessing, distance calculations are performed using the \emph{distance\_matrix} \rev{and distance.cdist function of the module \emph{scipy.spatial}, }and random selection uses \emph{random package}. Graphical illustrations are generated using \emph{matplotlib} and \emph{seaborn}. CASTRO is available under the GNU GPL v3.0 license. Additional information can be found on the GitHub page \citep{githubcastro}. The data that supports the findings presented in \Cref{sec:res} are also available in the CASTRO GitHub repository at \url{https://github.com/AMDatIMDEA/castro/tree/main/examples/data}.
	
	\section{Results}\label{sec:res}
	\subsection{Four Dimensional Material Composition Problem}\label{sec:ex1}
	Consider a scenario in which a chemist needs to identify additional experiments to perform within a limited budget. The goal is to fully explore the design space, with a budget fixed at 15 experiments. Previously, 75 experiments have been conducted and these must be taken into account in the exploration. The four components under investigation are biobased polyamide (PA-56), phytic acid (PhA), an amino-based component, and a metal-containing component.
	The chemist will choose the specific amino and metal-containing components. The bounds are set as follows 
	\begin{align}
		&0.8 \leq \text{PA-56} \leq 1,\\
		&0 \leq \text{PhA} \leq 0.05,\\
		&0 \leq \text{amino-based component} \leq 0.1,\\
		&0 \leq \text{metal-containing component} \leq 0.14.
		\notag
	\end{align}
	The fractions of all components need to sum up to 1, i.e.
	\begin{equation}
		\text{PA-56}+\text{PhA}+ \text{amino-based component}+ \text{metallic-based component}=1.
	\end{equation}
	We begin with a total of 144 samples. Considering the 4 factorial permutations of the bounds, we sample six points for each permutation and select all feasible samples according to the algorithm. \rev{We use a stochastic version of the algorithm, running it with 5 different random seeds. From the results, we randomly select the minimum number of samples across the runs and combine them.} This process yields \rev{$97\times 5$} feasible samples for the LHS variant and \rev{$95\times5$} feasible samples for the LHSMDU variant. \rev{We select the 90 samples that maximize uniformity by using pairwise Euclidean distances.}
	
\begin{figure}[htbp]
		\centering
		\includegraphics[width=\textwidth]{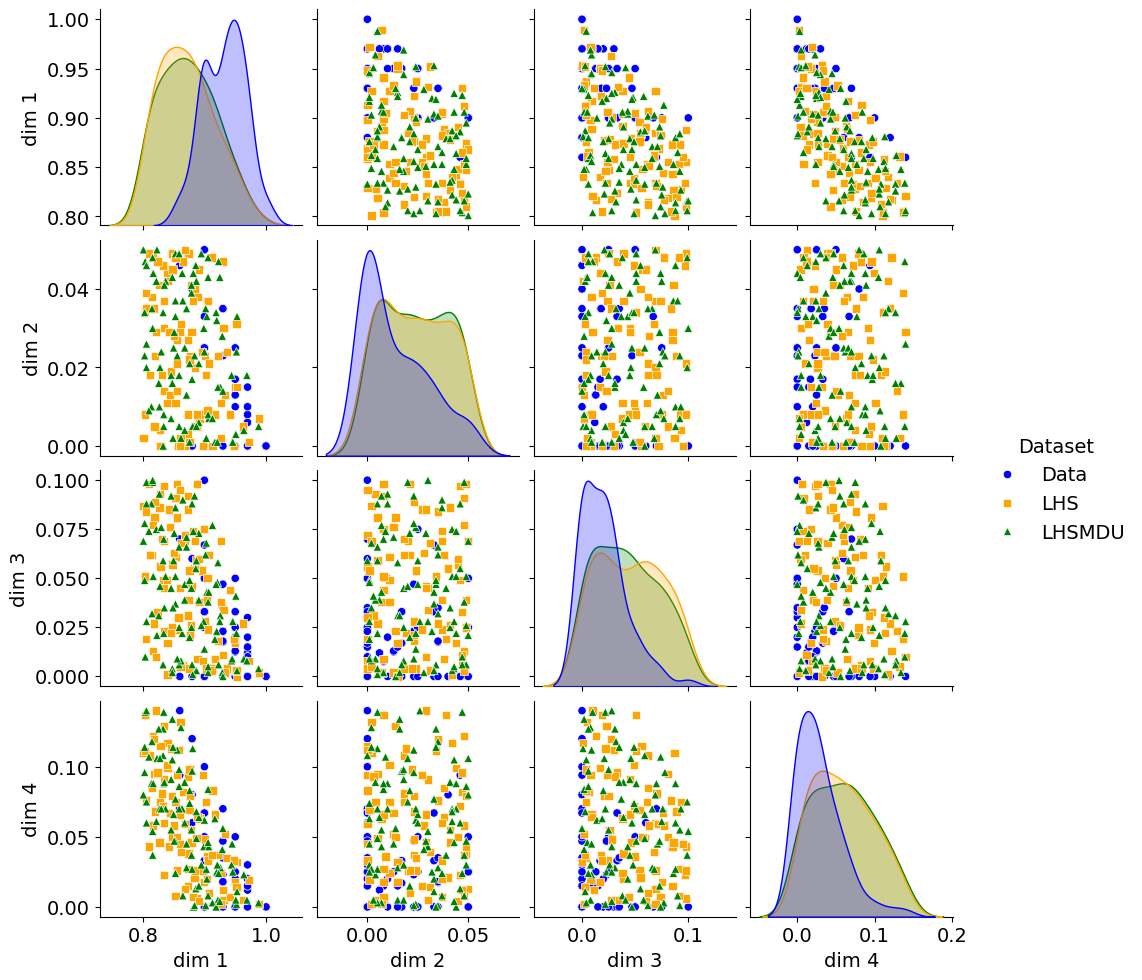}
		\caption{\rev{90 suggestions for the 4-dimensional problem. Dim 1,\ldots,4 corresponds to PA-56, PhA, the amino-based component and the metal-containing component respectively.}}
		\label{Fig:90sugg4dim}
	\end{figure}

\begin{figure}[htbp]
		\centering
		\includegraphics[width=\textwidth]{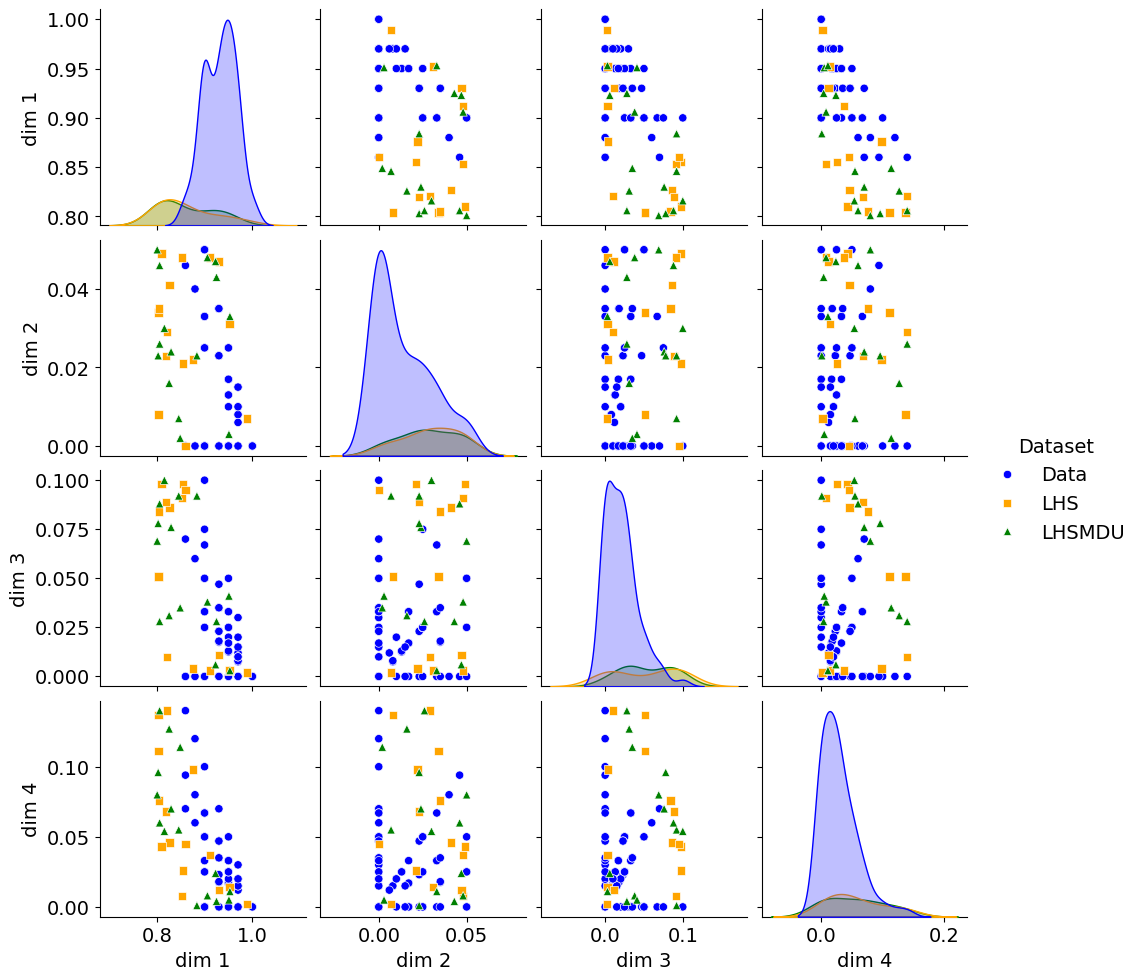}
		\caption{\rev{15 suggestions for the 4-dimensional problem. Dim 1,\ldots,4 corresponds to PA-56, PhA, the amino-based component and the metal-containing component respectively.}}
		\label{Fig:15sugg4dim}
	\end{figure}

\begin{figure}[htbp]
		\centering
		\includegraphics[width=\textwidth]{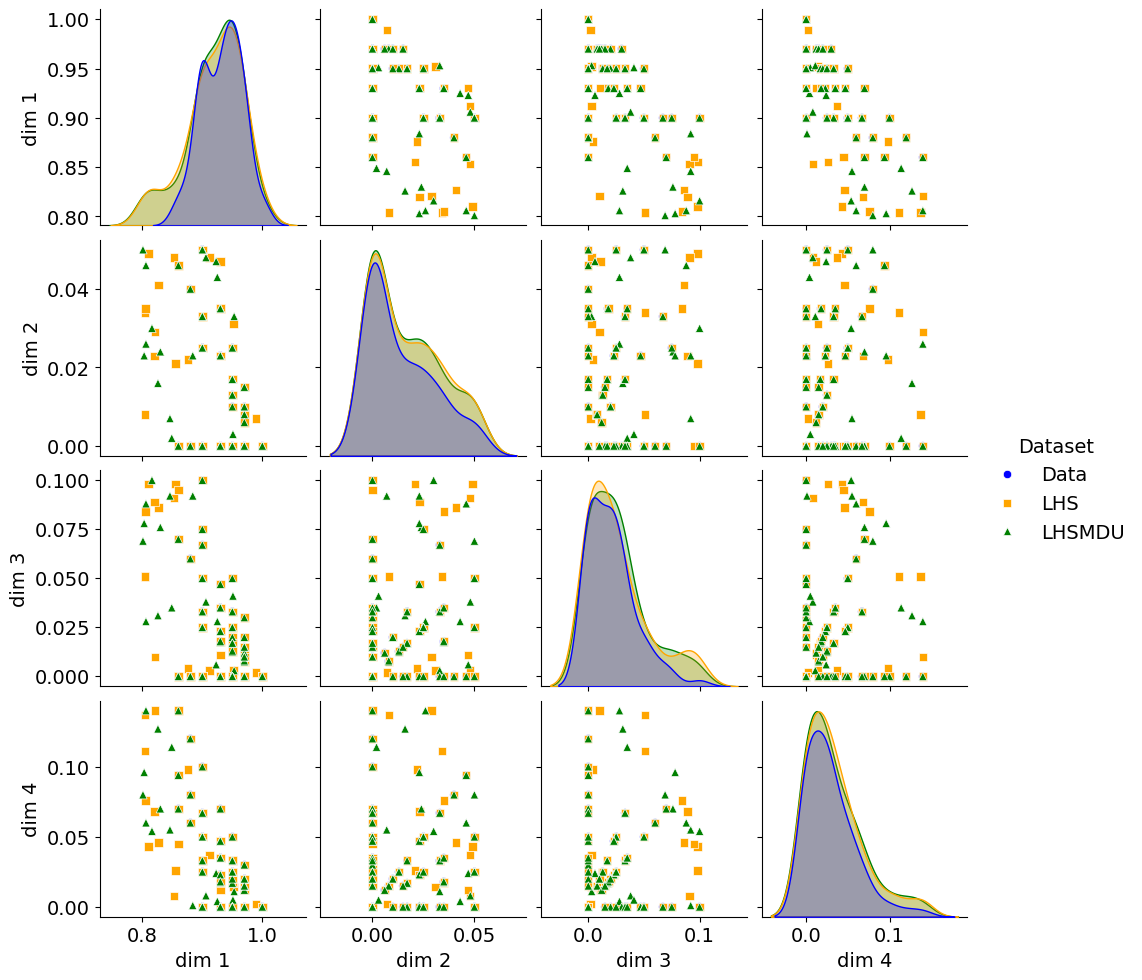}
		\caption{\rev{15 suggestions plus data for the 4-dimensional problem. Dim 1,\ldots,4 corresponds to PA-56, PhA, the amino-based component and the metal-containing component respectively.}}
		\label{Fig:15plusdatasugg4dim}
	\end{figure}

	\rev{The pairwise distributions of the 90 suggestions for all components, generated using CASTRO\textsubscript{LHS} and CASTRO\textsubscript{LHSMDU}, are compared to the previously collected data and illustrated in \cref{Fig:90sugg4dim}. Subsequently, a distance-based postprocessing step is applied to these 90 samples relative to the original data, reducing them to 15 experimental recommendations. The 15-point subsets derived from both algorithm variants are shown in \cref{Fig:15sugg4dim}, while \cref{Fig:15plusdatasugg4dim} presents the 15 points combined with the initial dataset. Notably, most CASTRO-generated points exhibit substantial deviation from the experimental data.}
	
	\rev{The distributions of the experimental data (blue circles) are biased, as illustrated in \cref{Fig:90sugg4dim}. In contrast, the CASTRO sampling methods generate distributions that approximate uniform coverage across the parameter space for the 90 samples. Among the two methods, CASTRO\textsubscript{LHSMDU} (green triangles) appears to provide better space coverage compared to CASTRO\textsubscript{LHS} (orange squares).}
	
	\rev{Furthermore, the CASTRO methods clearly extend sampling to areas that were underrepresented in the original experimental data. By addressing these previously unexplored regions, both methods contribute to a more comprehensive exploration of the parameter space, with CASTRO\textsubscript{LHSMDU} providing a more consistent and uniform coverage.} 
	
	The distributions for the remaining 15 suggestions, after removing those close to previously conducted experimental points, are shown in \rev{\cref{Fig:15sugg4dim}. Both CASTRO\textsubscript{LHS} (orange squares) and CASTRO\textsubscript{LHSMDU} (green triangles) help extend the coverage of the design space. In detail, CASTRO\textsubscript{LHS} explores regions outside the primary clusters of the Data, contributing new samples in underrepresented areas, albeit with some clustering.
		The LHSMDU variant, being more uniform by design, achieves even better distribution, filling gaps in the space that neither Data nor the LHS variant cover effectively. Clear differences are observed between the two variants. The first three components exhibit similar trends, but the last component is sampled closer to the lower bound for CASTRO\textsubscript{LHSMDU} and slightly closer to the upper bound for CASTRO\textsubscript{LHS}. The distributions of the combined set of 15 suggestions plus the data, as illustrated in \cref{Fig:15plusdatasugg4dim}, confirm the complementary roles of the two CASTRO methods. It should be noted that the blue Data points are here covered by the orange and green CASTRO points including the data since this figure highlights the 15 suggestions plus the previously collected data. 
		The combined plots demonstrate how CASTRO\textsubscript{LHS} and CASTRO\textsubscript{LHSMDU} effectively supplement the biased data distribution, enhancing the diversity and uniformity of the overall dataset.} 
	
	\rev{In addition to pairwise distribution analysis, we evaluated standard metrics, including central and wrap-around discrepancy as well as variance, to assess the space-filling properties of our new CASTRO designs. These metrics were compared against scaled traditional LHS and LHSMDU methods. For discrepancy calculations, we employed the \emph{scipy.stats.qmc.discrepancy} module. The scaled methods involve applying traditional LHS or LHSMDU, respectively, and then scaling the results to conform to the inequality bounds, ensuring that the components of each sample sum to 1. Sampling was conducted using the same five seeds as for our methods, \revt{here selecting the 90 samples with the largest pairwise Euclidean distances to maximize uniformity for fair comparison.} The resulting metrics from the method comparison are summarized in \cref{Tab:disc4dim}.}
	
	\revt{In particular, the central discrepancy (CD) metric was used to evaluate the uniformity of the design points in the central region of the space, with lower values indicating a more even spread of points. For the wrap-around discrepancy (WD), we assessed the distribution of points at the boundaries, where lower values indicate a more uniform coverage of the space's edges. Both CD and WD are critical in ensuring that the design does not favor certain regions of the space while neglecting others, thus achieving better overall space-filling properties.}
	
	\revt{We also analyzed variance, which measures the overall dispersion of the design points across the space. Higher variance can indicate greater flexibility and coverage across the space, though it can also reduce consistency if not balanced correctly. For a design to achieve optimal space-filling properties, it is important to strike a balance between lower discrepancy (for uniformity) and controlled variance (for flexibility and coverage).} 
	
	
	\begin{table}[htbp]
		\centering
		\begin{tabular}{|c|c|c|c|c|}
			Method&\# pts&CD&WD&Var\\
			\hline
			\multirow{3}{*}{CASTRO\textsubscript{LHS}} & \cellcolor{subrowcolor1} 15 & \cellcolor{subrowcolor1}  0.1638& \cellcolor{subrowcolor1}  0.2414& \cellcolor{subrowcolor1} 0.1212 \\
			\cline{2-5}
			& \cellcolor{subrowcolor2} 15 + data & \cellcolor{subrowcolor2}  0.3376& \cellcolor{subrowcolor2}  0.2600& \cellcolor{subrowcolor2} 0.0975\\
			\cline{2-5}
			& \cellcolor{subrowcolor3} 90 & \cellcolor{subrowcolor3}  0.0537& \cellcolor{subrowcolor3} 0.0541 & \cellcolor{subrowcolor3}  0.0821\\
			\hline
			\multirow{3}{*}{LHS\textsubscript{scaled}} & \cellcolor{subrowcolor1} 15 & \cellcolor{subrowcolor1}  0.3038& \cellcolor{subrowcolor1} 0.2970 & \cellcolor{subrowcolor1}0.0868 \\
			\cline{2-5}
			& \cellcolor{subrowcolor2} 15 + data & \cellcolor{subrowcolor2}  0.4284& \cellcolor{subrowcolor2}  0.3087& \cellcolor{subrowcolor2}  0.0901\\
			\cline{2-5}
			& \cellcolor{subrowcolor3} 90 & \cellcolor{subrowcolor3}  0.2941& \cellcolor{subrowcolor3}  0.3172& \cellcolor{subrowcolor3}  0.0556\\
			\hline
			\multirow{3}{*}{CASTRO\textsubscript{LHSMDU}} & \cellcolor{subrowcolor1} 15 & \cellcolor{subrowcolor1} 0.1129 & \cellcolor{subrowcolor1}  0.1221& \cellcolor{subrowcolor1}  0.1061\\
			\cline{2-5}
			& \cellcolor{subrowcolor2} 15 + data & \cellcolor{subrowcolor2} 0.3352& \cellcolor{subrowcolor2} 0.2528& \cellcolor{subrowcolor2}  0.0948\\
			\cline{2-5}
			& \cellcolor{subrowcolor3} 90 & \cellcolor{subrowcolor3}  0.0517& \cellcolor{subrowcolor3} 0.0466& \cellcolor{subrowcolor3} 0.0817\\
			\hline
			\multirow{3}{*}{LHSMDU\textsubscript{scaled}} & \cellcolor{subrowcolor1} 15 & \cellcolor{subrowcolor1}  0.2282& \cellcolor{subrowcolor1} 0.2855& \cellcolor{subrowcolor1}  0.0668\\
			\cline{2-5}
			& \cellcolor{subrowcolor2} 15 + data & \cellcolor{subrowcolor2} 0.4135 & \cellcolor{subrowcolor2}  0.3000& \cellcolor{subrowcolor2} 0.0867\\
			\cline{2-5}
			& \cellcolor{subrowcolor3} 90 & \cellcolor{subrowcolor3}  0.2751& \cellcolor{subrowcolor3} 0.3220 & \cellcolor{subrowcolor3}  0.0512\\
			\hline
		\end{tabular}
		\caption{\revt{Comparison of discrepancy (Central=CD and Warp-around=WD) and variance for CASTRO and scaled LHS/LHSMDU (with mixture constraint but not synthesis constraint, just theoretical baseline) for the 4-dimensional problem.}}
		\label{Tab:disc4dim}
	\end{table}
	
	\begin{figure}[htbp]
		\centering
		\includegraphics[width=\textwidth]{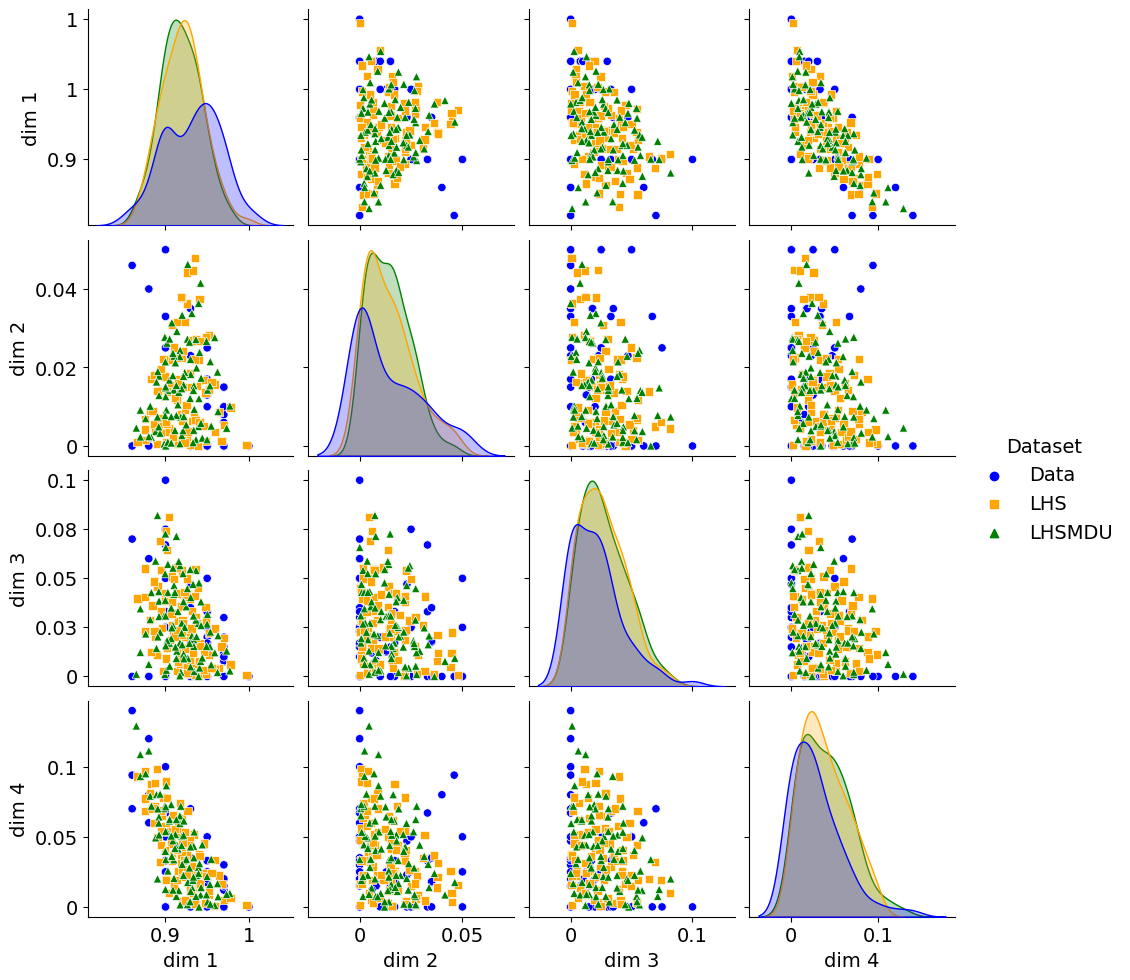}
		\caption{\revt{90 scaled traditional LHS/LHSMDU theoretical baseline suggestions for the 9-dimensional problem. Dim 1,\ldots,4 corresponds to PA-56, PhA, the amino-based components, and the metal-containing components.}}
		\label{Fig:90sugg4dimLHSLHSMDU}
	\end{figure}
	
	\rev{CASTRO\textsubscript{LHS} and CASTRO\textsubscript{LHSMDU} show lower discrepancy than scaled LHS and LHSMDU, respectively, for 15 points, 15 points plus the experimental data, and 90 points. Additionally, CASTRO exhibits higher variance for all cases except the 90 points, likely due to the distance check that filters out these points.}
	
	\revt{Discrepancy measures how evenly points are distributed across the design space. A lower discrepancy indicates better coverage across both the central and boundary regions, ensuring efficient exploration. CASTRO's lower discrepancy suggests more even space filling compared to scaled LHS and LHSMDU, without clustering in any region.}
	
	\revt{Variance reflects the spread or consistency of the points. Higher variance can mean more dispersion, which may seem less stable but is acceptable for ensuring thorough coverage of boundary regions. CASTRO's higher variance reflects its flexibility in covering diverse areas of the space, preventing over-concentration. However, for the 90-point case, CASTRO shows lower variance, indicating more stability while maintaining effective coverage.}
	
	\revt{When comparing the distributions of the 90 CASTRO suggestions (\cref{Fig:90sugg4dim}) with the 90 scaled traditional LHS/LHSMDU suggestions (\cref{Fig:90sugg4dimLHSLHSMDU}), clear differences emerge in terms of space coverage and clustering tendencies. The pairwise distribution plots for scaled traditional LHS/LHSMDU for 15 points and 15 points plus data are provided in the Additional Figures section of the Supplementary Material \citep{supplementary}.}
	
	\revt{The traditional methods, particularly LHS (orange squares), exhibit noticeably more clustering, especially in the center of dim 1 and within the lower to medium value ranges across the other dimensions. In the marginal plots, LHS distributions extend beyond the original Data (blue circles), broadening coverage, but doing so unevenly.}
	
	\revt{LHSMDU (green triangles) performs better in terms of uniformity, exploring boundary regions that both Data and LHS tend to underrepresent. This is particularly evident in dimensions like dim 3 and dim 4, where LHSMDU fills gaps closer to the extremes of the design space. Despite this improvement, both LHS and LHSMDU still exhibit localized clustering patterns and do not completely eliminate the gaps in the design space.}
	
	\revt{In contrast, CASTRO demonstrates a more balanced approach, minimizing clustering while maximizing coverage across the entire design space. Its sampling strategy not only introduces new samples in underrepresented regions, but does so more efficiently, ensuring that central and boundary areas are explored without unnecessary redundancy. Therefore, based on the distribution analysis, discrepancy, and variance metrics, CASTRO methods are better suited for constrained design spaces since they strike a better balance between coverage and distribution.}
	
	\revt{In this example, CASTRO\textsubscript{LHS} and CASTRO\textsubscript{LHSMDU} show very similar discrepancies for the 15 points plus data, indicating similarly even space coverage across the center and boundaries. However, for both the 15-point and 90-point designs, CASTRO\textsubscript{LHSMDU} outperforms CASTRO\textsubscript{LHS}, with lower discrepancy and similar variance. Based on the illustrations in \cref{Fig:90sugg4dim,Fig:15sugg4dim,Fig:15plusdatasugg4dim}, it appears that the LHSMDU variant provides more complementarity to the original data. Therefore, for this 4-dimensional problem, we recommend that the chemist conduct the next 15 experiments based on the CASTRO\textsubscript{LHSMDU} points.}
	
	\subsection{Nine Dimensional Material Composition Problem}\label{sec:ex2}
	The chemist now seeks our assistance in identifying additional experiments to perform within a limited budget while specifying all components, rather than selecting from certain categories based on experience. This task involves the transition from a simple four-dimensional problem of material composition to a more complex nine-dimensional problem, which requires additional steps outlined in \cref{sec:meth}. 
	
	To address this challenge, we divide the nine-dimensional problem into three subproblems. The primary problem remains as described in \Cref{sec:ex1}. Additionally, we create two subproblems: one focusing on four amino-based components—Chitosan (CS), Boron Nitride (BN), Tromethamine (THAM), and Melamine (MEL)—and another centered on three metal-containing components—Calcium Borate (CaBO), Zinc Borate (ZnBO), and Halloysite Nanotube (HNT). 
	
	For the original problem and thus, the principal problem and 2 subproblems the components' fractions need to sum up to 1.
	
	In line with the previous scenario, conducting experiments remains costly, and our budget limits us to performing only a certain number of new experiments in addition to the existing database of 75 samples. For this nine-dimensional problem, we are restricted to conducting 15 new experiments. We aim to thoroughly explore this expanded design space while considering the constraints of our budget and the data from previous experiments.
	
	To address this task, we apply the CASTRO algorithm to all three subproblems and then integrate the results. First, we sample with CASTRO for subproblem 1, represented by the problem from the \rev{previous \cref{sec:ex1}.} We stop when we receive the 90 CASTRO\textsubscript{LHS} and \linebreak 90 CASTRO\textsubscript{LHSMDU} suggestions for this subproblem. 
	
	We continue with subproblem 2, i.e. the amino-based problem, and we begin with a total of \rev{384} samples. Sampling \rev{16} points per permutation of the bounds, we select all feasible samples obtained through the algorithm.  \rev{Using the stochastic version of the algorithm, we sample \rev{16} points per permutation of the bounds across 5 different random seeds. We combine the resulting samples by randomly selecting the minimum number of samples across the runs.} 
	This process yields \rev{$99\times5$} feasible samples for the LHS variant and \rev{$101\times5$} feasible samples for the LHSMDU variant. \rev{To maximize uniformity, we select the 90 samples with the largest pairwise Euclidean distances.} 
	
	\rev{This problem exhibits a higher rejection rate due to its looser bounds, thus, significantly increasing the difficulty. Each component has a lower bound of 0 and an upper bound of 1, i.e. 
		\begin{align}
			&0 \leq \text{CS} \leq 1,\\
			&0 \leq \text{BN} \leq 1,\\
			&0 \leq \text{THAM} \leq 1,\\
			&0 \leq \text{MEL} \leq 1,\\
			\notag
		\end{align}
		creating a larger feasible region compared to subproblem 1. These expansive bounds increase the likelihood of generating infeasible samples, making the selection process more challenging.}
	The fractions of all amino-based components need to sum up to 1, i.e.
	\begin{equation}
		\text{CS}+\text{BN}+ \text{THAM}+ \text{MEL}=1.
	\end{equation}
	After sampling using CASTRO for the amino-based problem, we ensure that the final combinations can be synthesized. Only specific combinations are allowed, such as Mel+CS, THAM+CS, and Mel+THAM, while Mel, THAM, CS, and BN are also permissible as single amino components. To handle these synthesis restrictions, we introduce additional mixture constraints. For the single component constraints this translates into integer constraints where instead of directly considering integer variables, we treat them as real variables and then in the post-processing stage, employ rounding strategies for integer transformation. We select combinations where the fraction, that is, $comp^i_k$, $k=1,...n_{comp}$, $i=1,...n_{feas}$ for component $k$ and sample $i$ is greater than 0.5, that is, $comp^i_k\ge0.5$. $n_{comp}$ denote the number of components and $n_{feas}$ the feasible CASTRO samples. If no valid combination with the second-largest value was selected in CASTRO, we round the fraction to 1. Alternatively, we choose the valid combination with the second-largest value, ensuring that their fractions sum to one. Finally, we randomly select 90 points from all feasible post-processed points.
	
	Furthermore, similar to the amino-based problem, for subproblem 3, the metal-based problem, we set the following bounds for all components: 
	\begin{align}
		&0 \leq \text{CaBO} \leq 1,\\
		&0 \leq \text{ZnBO} \leq 1,\\
		&0 \leq \text{HNT} \leq 1.\\
		\notag
	\end{align}
	As for the previous problem, the fractions of all metallic-based components need to sum up to 1, i.e.
	\begin{equation}
		\text{CaBO}+\text{ZnBO}+ \text{HNT}=1.
	\end{equation}
	Starting with 120 initial total samples for the three factorial permutations, we sample 20 points per permutation \rev{over 5 random seeds as for the previous problems. The results are combined by randomly selecting the minimum number of samples across the runs, leading to $94\times5$ feasible CASTRO\textsubscript{LHS} samples and $93\times5$ feasible CASTRO\textsubscript{LHSMDU} samples. Among each of these set of samples we select the 90 samples with the largest pairwise Euclidean distances to maximize uniformity. Additionally, we have to ensure an additional synthesis constraint that dictates that no combinations between metal-containing components are allowed, i.e. $comp_k^i \in \{0,1\}\forall k,i$.}
	
	To address this integer constraint, we post-process the selected CASTRO points by setting the component with the maximum fraction $comp_k^i$ to one and all others to zero, i.e. 
	\begin{equation}
		\begin{cases}
			&1, \quad\text{if} \quad k = \argmax_k \; comp_k^i, \;k=1,...n_{comp}\\
			&0, \quad\text{else}.
		\end{cases} 
	\end{equation}
	$\forall i=1,...n_{feas}$.
	This adjustment ensures that the fractions sum up to one again. From these feasible CASTRO points, we randomly select 90 points.
	
\begin{figure}[htbp]
		\begin{subfigure}[c]{0.5\textwidth}\centering\vspace{.12em}\includegraphics[scale=0.4]{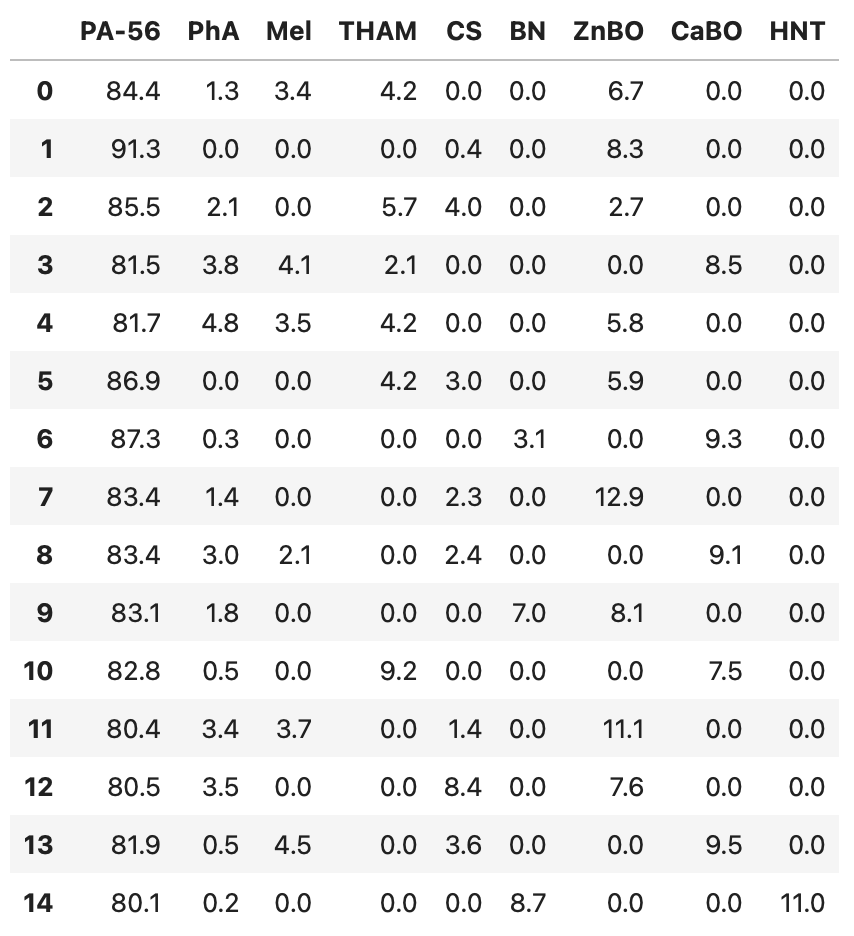}
			\caption{CASTRO\textsubscript{LHS} suggestions.}
		\end{subfigure}\begin{subfigure}[c]{0.5\textwidth}    \centering\includegraphics[scale=0.4]{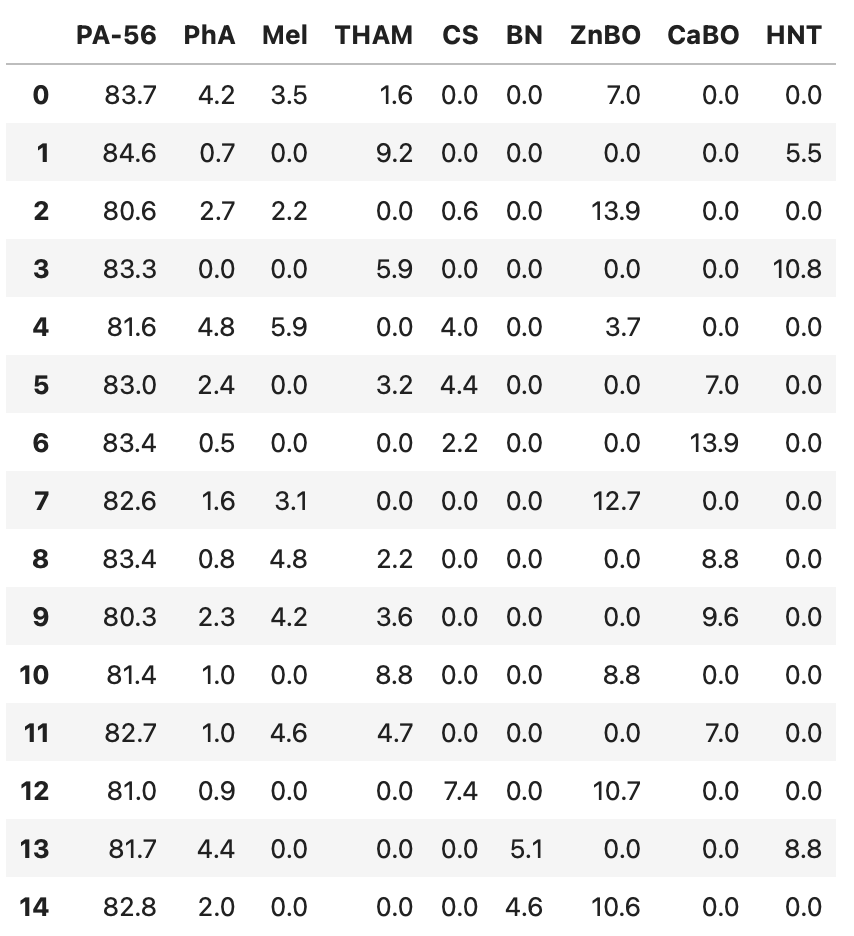}
			\caption{CASTRO\textsubscript{LHSMDU suggestions}}
		\end{subfigure}
		\caption{\rev{15 resulting CASTRO\textsubscript{LHS} and CASTRO\textsubscript{LHSMDU} suggestions  (rounded to 3 digits and converted into \%).}}
		\label{fig:table_lhs_lhsmdu}
	\end{figure}

	Following this, we integrate the three problems back into the nine-dimensional problem. We then choose the 15 points with the farthest Euclidean distance from the previously collected data. The 15 CASTRO\textsubscript{LHS} and CASTRO\textsubscript{LHSMDU} suggestions obtained can be found in \cref{fig:table_lhs_lhsmdu} (rounded to 3 digits and converted into \%).
	
	\begin{figure}[htbp]
		\centering
		\includegraphics[width=\textwidth]{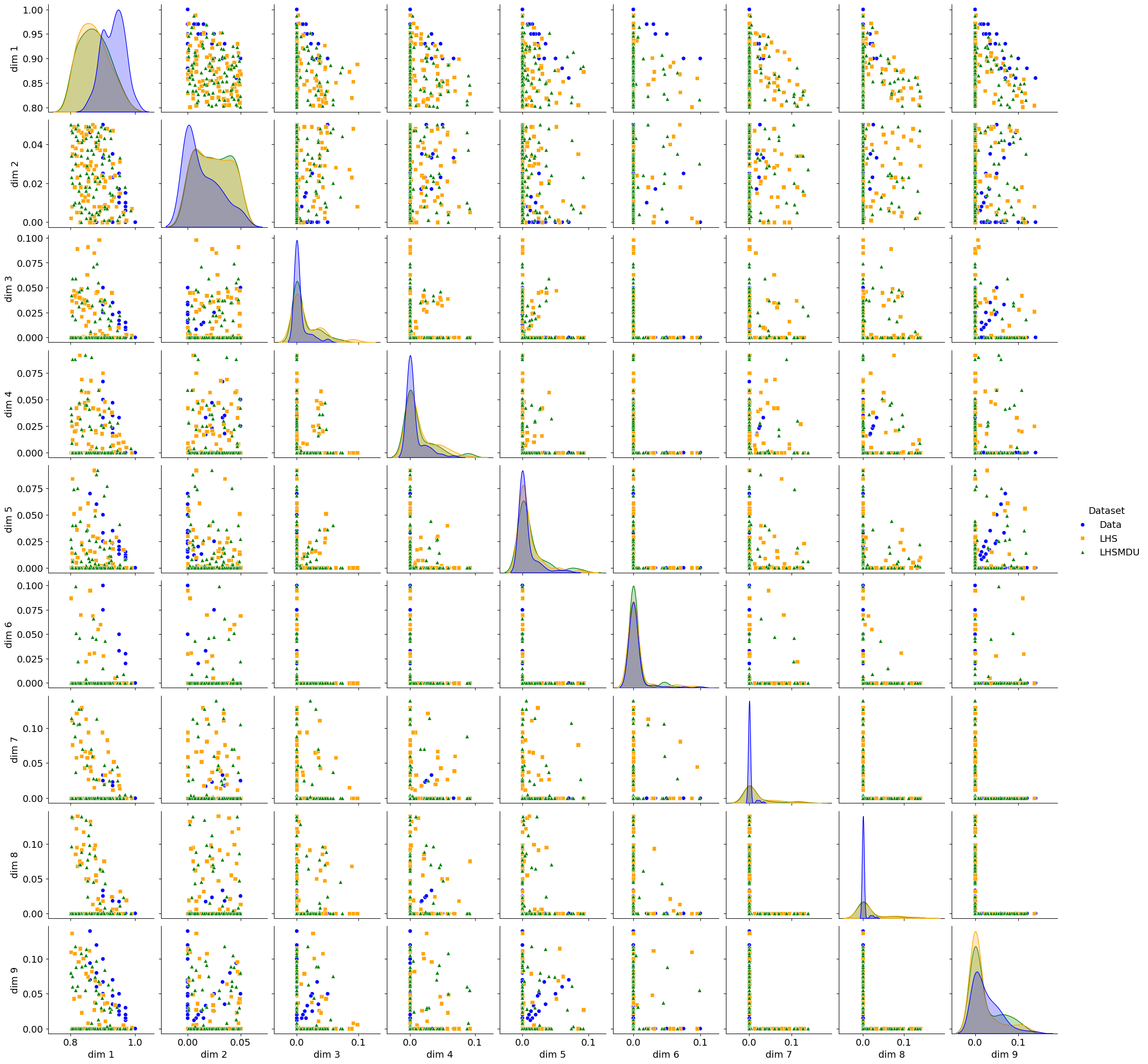}
		\caption{\rev{90 suggestions for the 9-dimensional problem. Dim 1,\ldots,9 corresponds to PA-56, PhA, the amino-based components, i.e. CS, BN, THAM, and MEL, and the metal-containing components, i.e. CaBO, ZnBO, and HNT respectively.}}
		\label{Fig:90sugg9dim}
	\end{figure}
	\begin{figure}[htbp]
		\centering
		\includegraphics[width=\textwidth]{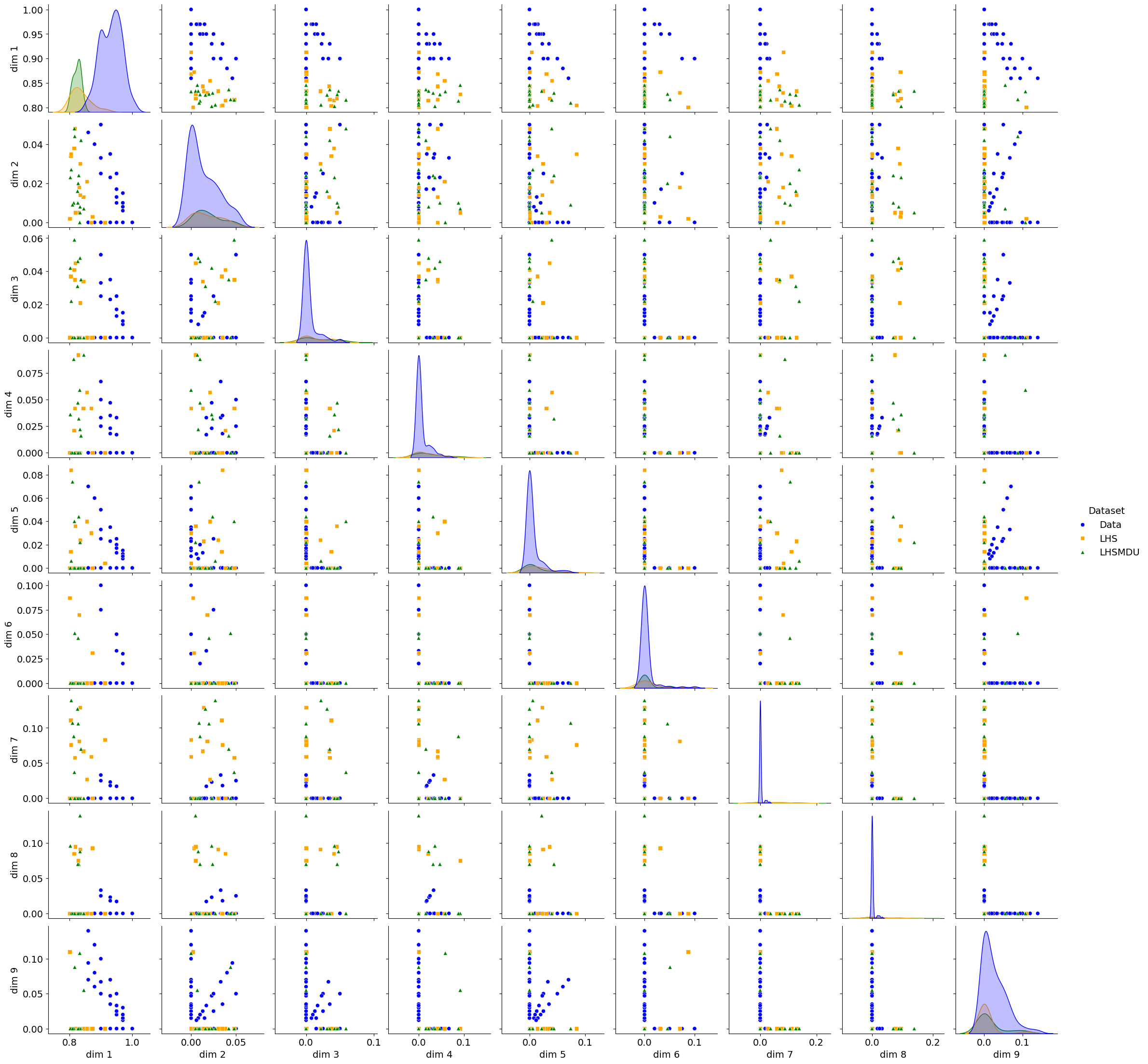}
		\caption{\rev{15 suggestions for the 9-dimensional problem. Dim 1,\ldots,9 corresponds to PA-56, PhA, the amino-based components, i.e. CS, BN, THAM, and MEL, and the metal-containing components, i.e. CaBO, ZnBO, and HNT respectively.}}
		\label{Fig:15sugg9dim}
	\end{figure}
	\begin{figure}[htbp]
		\centering
		\includegraphics[width=\textwidth]{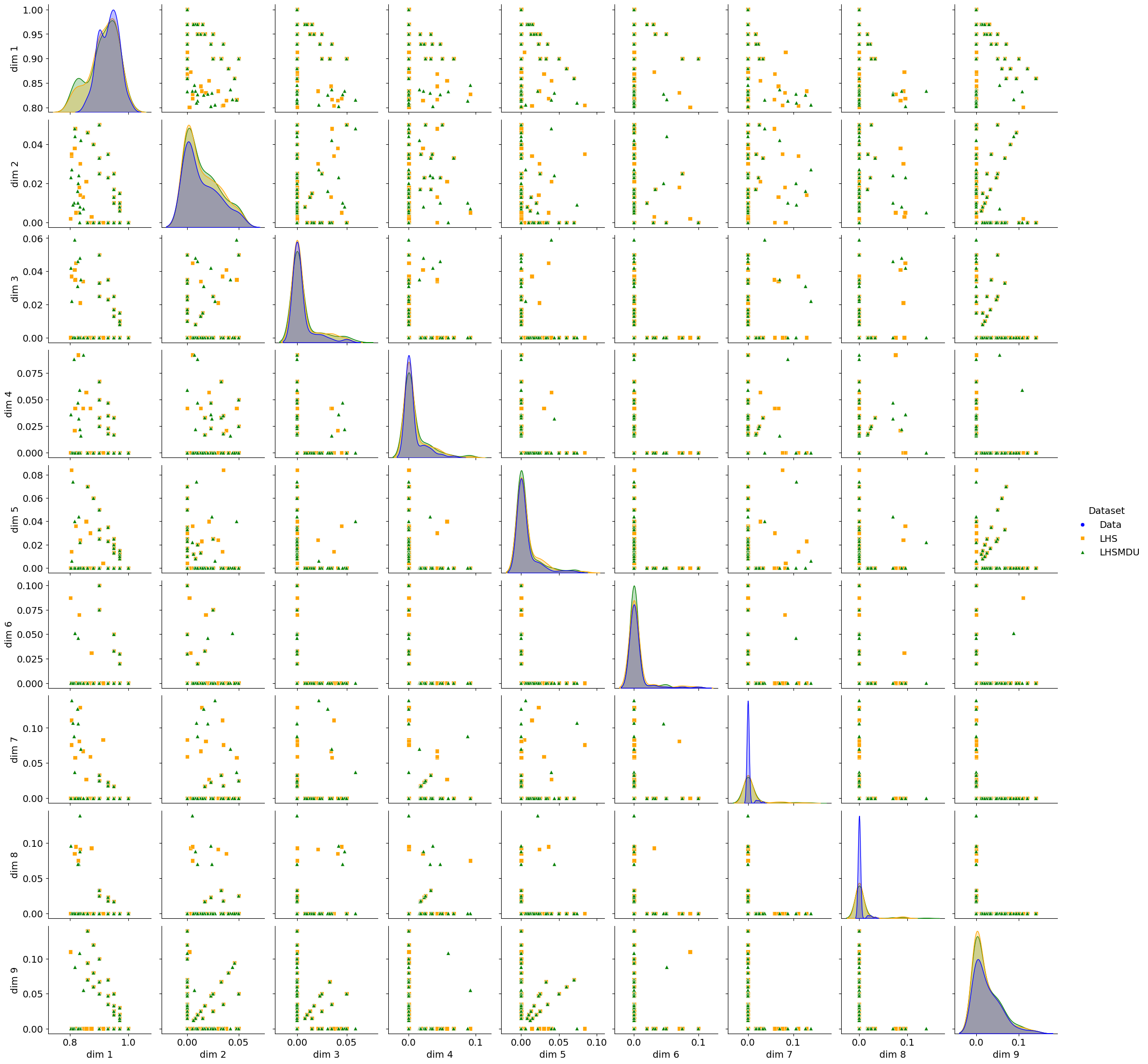}
		\caption{\rev{15 suggestions plus data for the 9-dimensional problem. Dim 1,\ldots,9 corresponds to PA-56, PhA, the amino-based components, i.e. CS, BN, THAM, and MEL, and the metal-containing components, i.e. CaBO, ZnBO, and HNT respectively.}}
		\label{Fig:15plusdatasugg9dim}
	\end{figure}
	
	\rev{The pairwise distributions for the 90 suggestions are shown in \cref{Fig:90sugg9dim}. The experimental data (blue circles) reveal clustering and gaps, suggesting that it may not uniformly cover the parameter space. CASTRO\textsubscript{LHS} (orange squares) offers better coverage than the experimental data, with fewer clusters and improved uniformity. CASTRO\textsubscript{LHSMDU} (green triangles) achieves the most evenly distributed points, showing minimal clustering and better space coverage.}
	
	\rev{In certain dimension pairs (e.g., dim 2 vs. dim 4 or dim 5 vs. dim 9), the experimental data exhibits visible gaps and clusters, highlighting poor coverage. Both CASTRO\textsubscript{LHS} and CASTRO\textsubscript{LHSMDU} address this issue, with CASTRO\textsubscript{LHSMDU} providing the most uniform distribution across the space.}
	
	\rev{Dimensions 3 to 9, which correspond to the amino-based and metallic-based components, involve additional synthesis constraints. This is reflected in the figure, where the experimental data (blue circles) show increased clustering and sparsity in these dimensions. For instance, in dimensions like dim 5 and dim 7, feasible regions are underrepresented, leaving noticeable gaps. While CASTRO\textsubscript{LHS} improves coverage in dimensions 3 to 9, slight clustering or unevenness remains in some regions (e.g. dim 3 vs. dim 8 or dim 6 vs. dim 9).}
	
	\rev{In contrast, the CASTRO\textsubscript{LHSMDU} dataset demonstrates the best performance in dimensions 3 to 9. It effectively balances the constraints while ensuring uniform coverage as allowed under the synthesis constraints. This is evident from the relatively even spread of points across the scatterplots for these dimensions. Compared to both the experimental data and CASTRO\textsubscript{LHS}, CASTRO\textsubscript{LHSMDU} more effectively explores feasible constrained regions. The scatterplots highlight that CASTRO\textsubscript{LHSMDU} excels in maintaining coverage, even under stringent synthesis constraints.}
	
	\rev{These findings are confirmed by the 15 most distant points from the data, as shown in \cref{Fig:15sugg9dim}. CASTRO\textsubscript{LHSMDU} significantly improves uniformity and fills gaps across both constrained and unconstrained dimensions. While CASTRO\textsubscript{LHS} also contributes positively, it is less effective than CASTRO\textsubscript{LHSMDU} in maintaining uniformity throughout the space. The distributions of the combined set of 15 suggestions and the data, shown in \cref{Fig:15plusdatasugg9dim}, confirm that the two CASTRO methods complement the Data. Note that as this figure highlights the 15 suggestions plus the previously collected data, the blue Data points are here covered by the orange and green CASTRO points that include the data. CASTRO\textsubscript{LHSMDU} provides the most uniform and comprehensive coverage, making it the best complement to the previously collected experimental data.}
	
	\begin{table}[htbp]
		\centering
		\begin{tabular}{|c|c|c|c|c|}
			Method&\# pts&CD&WD&Var\\
			\hline
			\multirow{3}{*}{CASTRO\textsubscript{LHS}} & \cellcolor{subrowcolor1} 15 & \cellcolor{subrowcolor1} 4.4657 & \cellcolor{subrowcolor1} 6.5500 & \cellcolor{subrowcolor1} 0.0719 \\
			\cline{2-5}
			& \cellcolor{subrowcolor2} 15 + data & \cellcolor{subrowcolor2} 8.8637 & \cellcolor{subrowcolor2} 9.5895 & \cellcolor{subrowcolor2} 0.0693\\
			\cline{2-5}
			& \cellcolor{subrowcolor3} 90 & \cellcolor{subrowcolor3} 5.0772 & \cellcolor{subrowcolor3} 6.5816 & \cellcolor{subrowcolor3} 0.0741 \\
			\hline
			\multirow{3}{*}{LHS\textsubscript{scaled}} & \cellcolor{subrowcolor1} 15 & \cellcolor{subrowcolor1} 6.9832 & \cellcolor{subrowcolor1} 9.7691 & \cellcolor{subrowcolor1} 0.0205 \\
			\cline{2-5}
			& \cellcolor{subrowcolor2} 15 + data & \cellcolor{subrowcolor2} 9.5457 & \cellcolor{subrowcolor2} 9.6411 & \cellcolor{subrowcolor2} 0.0604 \\
			\cline{2-5}
			& \cellcolor{subrowcolor3} 90 & \cellcolor{subrowcolor3} 7.2246 & \cellcolor{subrowcolor3} 10.2511 & \cellcolor{subrowcolor3} 0.0195 \\
			\hline
			\multirow{3}{*}{CASTRO\textsubscript{LHSMDU}} & \cellcolor{subrowcolor1} 15 & \cellcolor{subrowcolor1} 4.6616 & \cellcolor{subrowcolor1} 7.2182 & \cellcolor{subrowcolor1} 0.0793 \\
			\cline{2-5}
			& \cellcolor{subrowcolor2} 15 + data & \cellcolor{subrowcolor2} 8.8048 & \cellcolor{subrowcolor2} 9.7114 & \cellcolor{subrowcolor2} 0.0706 \\
			\cline{2-5}
			& \cellcolor{subrowcolor3} 90 & \cellcolor{subrowcolor3} 5.2415 & \cellcolor{subrowcolor3} 6.8488 & \cellcolor{subrowcolor3} 0.0751 \\
			\hline
			\multirow{3}{*}{LHSMDU\textsubscript{scaled}} & \cellcolor{subrowcolor1} 15 & \cellcolor{subrowcolor1}  6.6392& \cellcolor{subrowcolor1} 9.5679 & \cellcolor{subrowcolor1} 0.0178\\
			\cline{2-5}
			& \cellcolor{subrowcolor2} 15 + data & \cellcolor{subrowcolor2} 9.4326 & \cellcolor{subrowcolor2} 9.5611 & \cellcolor{subrowcolor2} 0.0600\\
			\cline{2-5}
			& \cellcolor{subrowcolor3} 90 & \cellcolor{subrowcolor3} 7.0710 & \cellcolor{subrowcolor3} 9.9767 & \cellcolor{subrowcolor3} 0.0195 \\
			\hline
		\end{tabular}
		\caption{\revt{Comparison of discrepancy (Central=CD and Warp-around=WD) and variance for CASTRO and scaled LHS/LHSMDU (with mixture but not synthesis constraints, just theoretical baseline) for 9-dimensional problem.}}
		\label{Tab:disc9dim}
	\end{table}
	
	As in the previous example, \revt{in addition to distribution analysis}, we assess the space-filling and uniformity of our resulting designs by evaluating the central and warp-around discrepancy, as well as the variance. This analysis is summarized in \cref{Tab:disc9dim}, where we compare the performance of CASTRO to the \rev{scaled} traditional LHS and LHSMDU methods as a theoretical baseline.\rev{This is due to the scaled traditional methods ensuring feasibility only concerning the mixture constraints, but not the here-present synthesis constraints.}
	\begin{figure}[htbp]
		\centering
		\includegraphics[width=\textwidth]{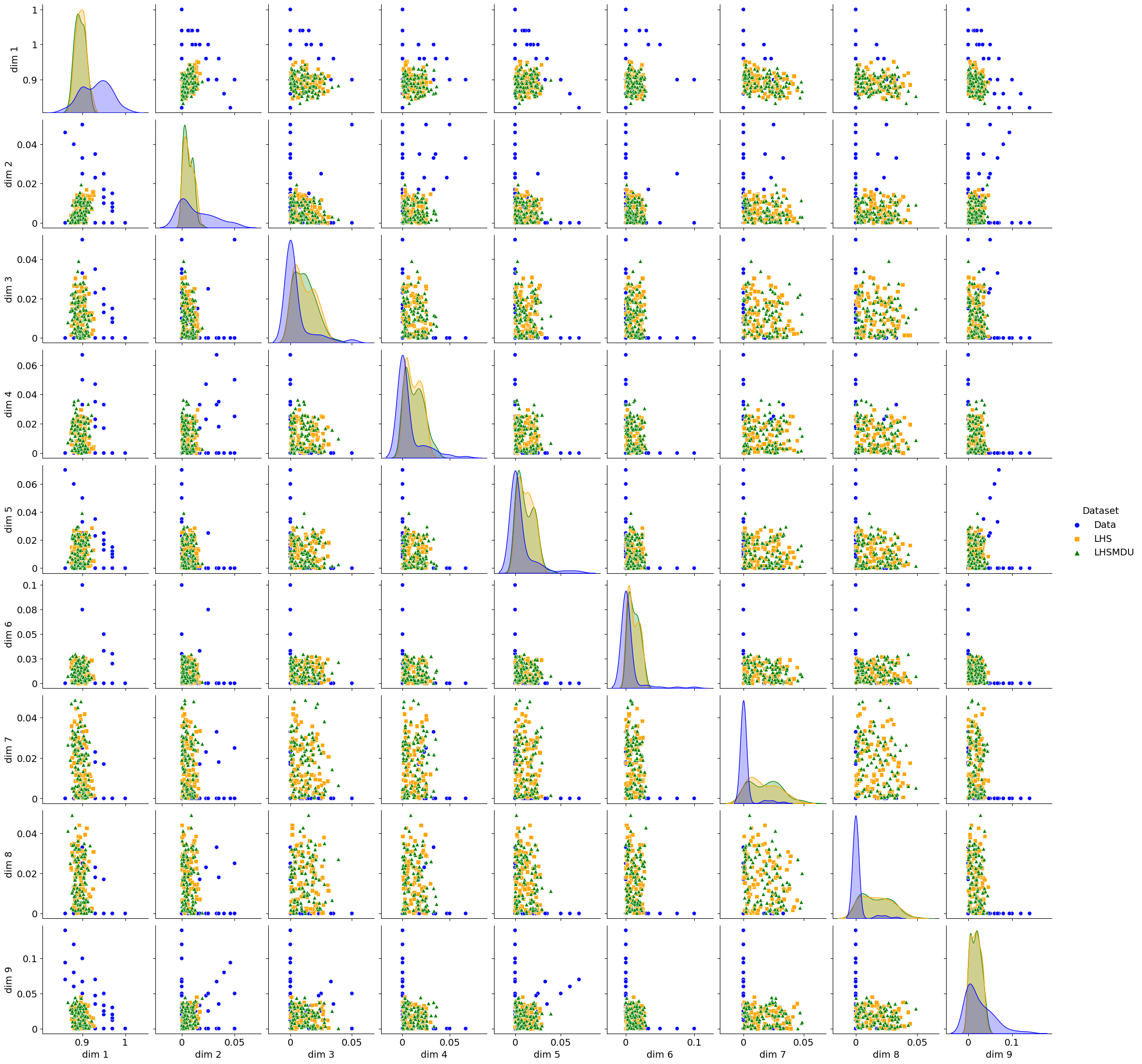}
		\caption{\revt{90 scaled traditional LHS/LHSMDU theoretical baseline suggestions for the 9-dimensional problem. Dim 1,\ldots,9 corresponds to PA-56, PhA, the amino-based components, i.e. CS, BN, THAM, and MEL, and the metal-containing components, i.e. CaBO, ZnBO, and HNT respectively.}}
		\label{Fig:90sugg9dimLHSLHSMDU}
	\end{figure}
	\revt{When comparing the distributions of CASTRO (\cref{Fig:90sugg9dim}) and scaled traditional LHS/LHSMDU (\cref{Fig:90sugg9dimLHSLHSMDU}) in the nine-dimensional case, clustering in the traditional LHS/LHSMDU becomes even more pronounced. Note that the pairwise distribution plots for scaled traditional LHS/LHSMDU for 15 points and 15 points plus data can be found in Additional Figures in Supplementary Material \citep{supplementary}.}
	
	\revt{For dim 2 through dim 9 (\cref{Fig:90sugg9dimLHSLHSMDU}), the samples are concentrated within approximately one-third of the permissible range. Specifically, the upper bounds are 0.05 for dim 2, 0.1 for dim 3 to dim 6, and 0.14 for dim 7 to dim 9. This heightened clustering is even more apparent than in the four-dimensional scenario, despite the absence of additional synthesis constraints in this analysis. While the traditional methods (orange squares, green triangles) broaden coverage in certain regions compared to the original data (blue circles), they fail to adequately sample near the upper boundaries. New samples are introduced into underrepresented regions, particularly in dim 3 to dim 8, but gaps remain in critical boundary areas. Nevertheless, LHSMDU (green triangles) achieves better coverage than LHS among the traditional methods.}
	
	\revt{CASTRO, in comparison, demonstrates better space-filling properties by effectively minimizing clustering and ensuring more uniform coverage across all nine dimensions under the additional synthesis constraints. Unlike traditional LHS and LHSMDU, CASTRO's sampling strategy efficiently explores both central regions and boundary extremes, addressing gaps that remain in the traditional methods. This is particularly evident in dimensions 3 through 8, where CASTRO introduces samples closer to the upper bounds and for dimension 1, where CASTRO introduces samples closer to the lower bound,  enhancing the diversity of the dataset. The CASTRO approach ensures that sampling is both comprehensive and balanced, offering a significant advantage in medium-dimensional experimental design problems with potential for high- dimensional problems.}
	
	\revt{Based on the distribution analysis, discrepancy, and variance metrics, CASTRO methods are preferable because they offer a better balance between coverage and distribution. For the 15-point, 15-point plus data, and 90-point designs, CASTRO\textsubscript{LHS} and CASTRO\textsubscript{LHSMDU} show lower or similar discrepancy compared to the traditional LHS and LHSMDU methods. This indicates that CASTRO methods more effectively cover the design space, leading to better overall point distribution. Note that the scaled traditional methods do not provide feasible solutions in this scenario because they do not account for the additional synthesis constraints, and thus serve only as theoretical baselines.}
	
	\revt{While CASTRO methods exhibit higher variance, this is generally an acceptable trade-off in experimental design, as the lower discrepancy suggests that the points are more evenly spread across the design space, which is crucial for achieving better results in practice. The higher variance can be seen as a reflection of the improved flexibility and coverage provided by CASTRO, compared to the traditional methods.}
	
	\revt{For the 15-point design, CASTRO\textsubscript{LHS} shows slightly lower discrepancy (CD and WD) and variance compared to CASTRO\textsubscript{LHSMDU}, indicating better coverage in the central region of the design space and at the boundaries but less variability. When combined with the data, CASTRO\textsubscript{LHS} results in a higher CD, indicating that the points are more spread out or scattered across the space, which can reduce uniformity but provides more extensive coverage. It also results in a lower WD, implying a more even distribution of points across the entire space, particularly around the boundaries. In addition, CASTRO\textsubscript{LHS} shows lower variance, suggesting greater stability in the design.}
	
	\revt{For the 90-point design, CASTRO\textsubscript{LHS} again shows slightly lower discrepancy (CD and WD) and variance than CASTRO\textsubscript{LHSMDU}. The lower variance here indicates a more consistent and stable spread of points, offering a balanced approach between maintaining uniformity (via low CD), good coverage of the edges (low WD) and reducing spread (via low variance). Based on these observations for the 15 points plus data and the visualizations, cf. \cref{Fig:90sugg9dim,Fig:15sugg9dim,Fig:15plusdatasugg9dim}, we recommend that the experimentalist use the 15 CASTRO\textsubscript{LHS} suggestions for the next experiments, as this will provide a design with balanced coverage (lower CD, WD) and variability (similar variance).}
	
	\section{Conclusion}\label{sec:con}
	In conclusion, this article introduces a novel methodology, available as the CASTRO software package, that enables sampling with equality mixture and other synthesis constraints while ensuring comprehensive space coverage within a limited budget. The method generates the desired number of feasible samples that cover the design space by effectively leveraging previously collected experimental data. It incorporates various techniques, including Latin hypercube sampling and Latin hypercube sampling with multidimensional uniformity. \rev{For problems }exceeding four dimensions, the method employs a divide-and-conquer strategy, breaking them down into more manageable subproblems.
	
	Upon introducing these new algorithms, we applied them to two material composition design examples: one with four dimensions and another with nine dimensions. In the case of the 4-dimensional problem, the method demonstrated distributions close to uniformity. However, the 9-dimensional problem introduced additional mixture constraints, resulting in specified distributions for most components.
	
	The novel method ensures space coverage through constrained sequential Latin hypercube sampling or Latin hypercube sampling with multidimensional uniformity. As a result, it provides a robust solution for experimental design, facilitating thorough exploration of the design space. Of particular significance is its applicability in scenarios with constrained budgets or prohibitively expensive experiments. The additional post-processing step of selecting samples farthest away from previously collected data points proves effective in addressing this challenge.
	
	Although the examples primarily focus on material composition design problems, the method's adaptability extends to various fields with similar constraints, such as the pharmaceutical and chemical industries. In essence, this methodology not only advances material science research but also offers promising solutions for addressing analogous challenges across diverse domains.
	
	Looking ahead, future work could explore extending the methodology to accommodate other types of constraints and incorporating additional sampling methods. \revt{While the current approach is optimized for small- to medium-dimensional problems, it is designed to be scalable. By leveraging the divide-and-conquer strategy, the method can be adapted to handle higher-dimensional problems, automating the division and parallel sampling of subproblems.} \revt{To further} \rev{enhance space coverage in high dimensions, a possible extension could involve exploring alternative distance metrics to the Euclidean distance, which may mitigate clustering around the mean in higher dimensions.}
	
	\section*{Acknowledgements}
	We are very grateful to José Hobson and De-Yi Wang (IMDEA Materials) for collecting the data that served for showcasing the new methodology. This article is part of the project TED2021-131409B-100, funded by MCIN/AEI/10.13039/501100011033 and by the European Union “NextGenerationEU”/PRTR.
	
	\section*{Supporting Information}
	\begin{enumerate}[label=\arabic*.] 
		\item \rev{Additional Figures}
		\begin{enumerate}[label=\arabic{enumi}.\arabic*.]
			\item \revt{Four-Dimensional Problem}
			\begin{enumerate}[label=\arabic{enumi}.\arabic{enumii}.\arabic*.]
				\item \revt{Scaled LHS and LHSMDU}
				\item \revt{CASTRO\textsubscript{LHS} and CASTRO\textsubscript{LHSMDU}}
			\end{enumerate}
			\item \revt{Nine-Dimensional Problem}
			\begin{enumerate}[label=\arabic{enumi}.\arabic{enumii}.\arabic*.]
				\item \revt{Scaled LHS and LHSMDU}
				\item \revt{CASTRO\textsubscript{LHS} and CASTRO\textsubscript{LHSMDU}}
			\end{enumerate}
		\end{enumerate}
		\item Algorithms
		\begin{enumerate}[label=\arabic{enumi}.\arabic*.]
			\item Algorithm 1: Conditioned Sampling Algorithm Dimension 1
			\item Algorithm 2: Conditioned Sampling Algorithm Dimension 2
			\item Algorithm 3: Conditioned Sampling Algorithm Dimension >2
			\item Algorithm 4: Permutation Subalgorithm
			\item Algorithm 5: Conditioned Sampling Subalgorithm Dimension >3
			\item Algorithm 6: Permutation Subalgorithm Dimension >3
		\end{enumerate}
	\end{enumerate}
	\section*{\rev{Declaration of generative AI and AI-assisted technologies in the writing process}}
	\rev{During the preparation of this work the authors used GPT-4 to assist in paraphrasing certain sections. After using this tool/service, the authors reviewed and edited the content as needed and take full responsibility for the content of the publication.}
	\bibliographystyle{elsarticle-harv}
	\bibliography{novelDOE_Bib}
\pagebreak
\begin{center}
	\textbf{\large Table of Contents Graphic}
\end{center}
\setcounter{equation}{0}
\setcounter{figure}{0}
\setcounter{table}{0}
\makeatletter
\renewcommand{\theequation}{S\arabic{equation}}
\renewcommand{\thefigure}{S\arabic{figure}}
\renewcommand{\bibnumfmt}[1]{[S#1]}
\renewcommand{\citenumfont}[1]{S#1}
\includegraphics[scale=0.45]{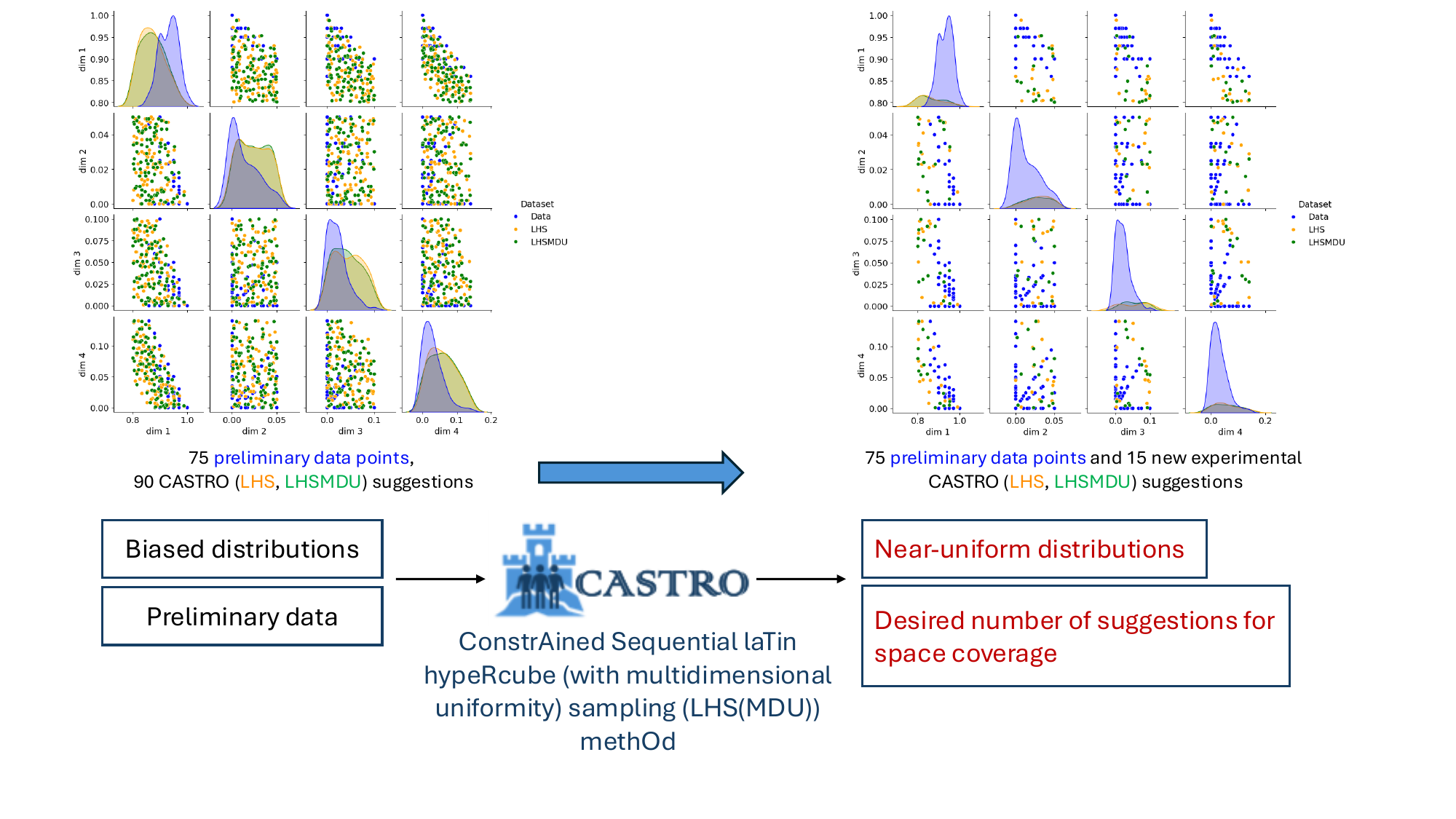}
\end{document}